\documentclass[pra,aps,groupedaddress,twocolumn,floatfix,nofootinbib]{revtex4-2}

\usepackage{hyperref}
\usepackage{multirow}
\usepackage{mathrsfs,amsmath} 
\usepackage{verbatim}
\usepackage{xcolor}
\usepackage{lipsum}
\usepackage{dsfont}
\usepackage{footnote}
\usepackage{amsfonts}
\usepackage{tcolorbox}
\usepackage{mathtools}
\usepackage{amsthm}
\usepackage{cleveref}
\usepackage{enumerate}
	
\usepackage[euler]{textgreek}
\definecolor{shadecolor}{rgb}{0.8,0.9,1}
\DeclareDocumentCommand{\Tr}{m m O{\big}}{{\rm Tr}_{\:\!{#1}}#3({#2}#3)}

\newcommand{\Q}{\mathbb{Q}}

\renewcommand{\nat}{\mathbb{N}}

\begin{document}
\title{Naturalistic intuitionism for physics}

\author{Bruno Bentzen}
\affiliation{School of Philosophy, Zhejiang University, 310058 Hangzhou, China}

\author{Flavio Del Santo}
\affiliation{Faculty of Physics, University of Vienna, 1090 Vienna, Austria;\\ and Group of Applied Physics, University of Geneva, 1211 Geneva 4, Switzerland}

\author{Nicolas Gisin}
\affiliation{Group of Applied Physics, University of Geneva, 1211 Geneva 4, Switzerland; and Constructor University, Bremen, Germany}

\date{\today}

\begin{abstract}
	Recently, a novel intuitionistic reconstruction of the foundations of physics has been primarily developed by Nicolas Gisin and Flavio Del Santo drawing on naturalism. Our goal in this paper is to examine and develop the philosophical background of their naturalistic intuitionism for physics in contrast with Brouwer's defense of his intuitionistic mathematics. To be exact, we propose a systematic rearticulation of Brouwer's so-called two acts of intuitionism to serve as the self-contained philosophical framework justifying naturalistic intuitionism in physics. This revision is accompanied by an investigation of the distinctive naturalistic treatment of some central intuitionistic topics, including logic, language, time, ontology, meaning, and truth.
\end{abstract}

\maketitle

%----------------------------------------------------------------------------------------------------
\section{Introduction} \label{intro}
%----------------------------------------------------------------------------------------------------

Over the last decades, there has been much discussion about whether constructive mathematics can provide the mathematical background necessary to do physics	\cite{hellman1993gleason,hellman1993constructive,bridges1995constructive,billinge1997constructive,richman2000gleason,fletcher2002constructivist}.
But it was not until the intuitionistic program put forward by Nicolas~Gisin~\cite{gisin2021indeterminism} in recent years that a form of constructivism began to be seriously advocated by physicists themselves. 
Motivated by the intrinsic randomness of quantum mechanics, its point of departure is the acceptance that physics need not presuppose a deterministic worldview. 
It therefore appears promising to abandon classical mathematics --- due to its inherently timeless nature --- in favor of intuitionistic mathematics, which provides a non-deterministic and temporal framework more suited to the foundations of physics. %glued together.
More precisely, the idea is to model indeterminism employing the intuitionistic notion of a choice sequence, namely, a potentially infinite sequence whose elements may be chosen freely or according to a law (algorithm). 

This novel intuitionistic program for physics has been developed by Nicolas Gisin and Flavio Del Santo mostly jointly in a sequence of papers~\citep{del2018striving, del2019physics, gisin2020real, gisin2021, del2021relativity, gisin2021indeterminism, del2021indeterminism, del2023prop, santo2023open, del2024creative, del2024features}.\footnote{
	See also~\citep{eyink2020renormalization,ben2020structure,dolev2024temporal} for further attempts in this direction.}  
In this paper, we develop the philosophical background that underlies this new program and examine it against the original defense of intuitionism by L. E. J. Brouwer~\cite{brouwer1975collected,brouwer1981cambridge}, thereby unveiling some crucial differences that until now have not been explicitly acknowledged in print. The main reason this is a valuable investigation to undertake is that the new intuitionism they advanced for physics is the product of a distinctive combination of naturalism and ontological realism, as opposed to the thorough idealism advocated by Brouwer without physical applications in mind. 
Indeed, as already admitted by Heyting, ``introspection is useless in physics''~\citep[p.~90]{heyting1974intuitionistic}. 
Yet, this naturalization of intuitionism shows that we can regardless retain some parts of intuitionistic mathematics and its philosophy that are useful in physics. 
It is therefore crucial that we identify the basic philosophical presuppositions that support this naturalization and how they differ from Brouwer's own intuitionistic views. 

%In fact, Heyting goes as far as to admit that classical mathematics is simpler to apply for mechanics, given that introspection is useless in physics~\citep{heyting1969mecanique,heyting1974intuitionistic}. 

To be exact, our goal here is to develop the philosophy of ``naturalistic intuitionism'' by means of a systematic revision of Brouwer's so-called two acts of intuitionism so as to serve as an autonomous naturalistic philosophical framework for physics. This development is then accompanied by an investigation of the distinctive naturalistic treatment of some central intuitionistic topics, including logic, language, time, ontology, meaning, and truth, that differ from Brouwer's understanding of these themes. All this is done in Section~\ref{sec:divergences}. 
The remaining sections of this paper are structured as follows. 
First, in Section~\ref{sec:nat-int}, we motivate and sketch the basic elements of naturalistic intuitionism in a self-contained way mainly targeted at physicists and philosophers of science as our intended audience, without assuming any previous familiarity with Brouwer's intuitionistic mathematics. 
Then, in Section~\ref{sec:brouwer}, we overview intuitionism as the original philosophy of mathematics put forward by Brouwer  through his so-called two acts of intuitionism. 
Section~\ref{sec:conclusion} provides some concluding remarks.

%----------------------------------------------------------------------------------------------------
\section{Naturalistic intuitionism} \label{sec:nat-int}
%----------------------------------------------------------------------------------------------------

Time is essential for physics because ``to be an event is to be in time''~\citep{dolev2018physics}. Without time we cannot tell stories and provide scientific explanations. %Without time we would not be able to tell stories, without which (scientific) explanations cannot~exist. 
Despite this, today's physics relegates time to a rather subaltern role. In most established physical theories, indeed, time is a mere parameter that labels a given succession of necessary events. And the situation is even more crystallized by a widespread relativistic worldview where time is geometrized to be just another coordinate,  akin to space. 
Quantum theory, in its standard interpretation, is an exception and, actually, is part of our motivation. 
With such a lean concept of time, physicists cannot speak of real evolutions nor surprises. More precisely, all apparent ``surprises'' are merely explained away by our ignorance: Turn a card face up, and we may be surprised by the figure that appears. The figure, however, has always been there, and the surprise is thus only due to a subjective lack of knowledge.
But real surprises, that is, events that are really --- in an ontological sense --- creating new information in the universe are hard or even impossible to describe with such a lean ``geometric'' view of time. In short, the current fundamental understanding of time in physics makes it hard to incorporate randomness, indeterminacy and creative evolutions. 

% \textcolor{red}{In short,  the current fundamental understanding of time in physics can only incorporate deterministic theories, its very structure allowing for no randomness, no indeterminacy, no creative evolution. \\COMMENT from Chris Wuthrich: \textit{eternalism is compatible wit indeterminism. Consider two hypothetical block universes (two possible worlds) which completely overlap in certain parts, but they differ in some other parts. This allows a block universe picture which is however indeterministic, because to the any state in the overlapping region of the two blocks correspond different continuations in the future and/or in the past.}
% \\ COMMENT from Flavio: \textit{I believe that explaining this here is unnecessary, to be honest. But also the way it is currently formulated is quite inaccurate. We should perhaps avoid the whole sentence.}}

However, what quantum theory strongly suggests is that some new events truly obtain: they happen although they were not predetermined. Such events are genuinely random; they involve the creation of new information—not by an agent, but existing in the universe, encoded in some physical medium—that did not exist prior to the occurrence of the event. One can, in fact, consider the fundamental distinction between ``before'' and ``after'' as a consequence of the emergence of novel information, thereby enabling the very notion of true becoming. 
We will return to this alternative conception of time later in \Cref{powerofnature}. 

Quantum physics supports the above-mentioned idea of indeterminism due to two main results: 
\begin{enumerate}[(i)]
	\item Heisenberg uncertainty relations (better-named indeterminacy relations);
	\item the violation of Bell inequalities. 
\end{enumerate}
Heisenberg relations impose an absolute minimal value (quantified by $\hbar$) to the smallest volume that a physical state can occupy, i.e., fundamental indeterminacy.\footnote{
    Note that indeterminacy is a kinematical feature, meaning that the pure state of a system does not determine all measurement outcomes. By contrast, indeterminism is a property of the dynamics: given the pure state (possibly of the universe) and the laws of physics, there need not be a unique continuation of the state into the future (nor into the past). Yet if measurements yield only a single outcome, indeterminacy leads to indeterminism, since a pure state leaves several outcomes open. From the same initial state, multiple future continuations are possible.}
However, this indeterminacy may exist only at the level of the theory and could, in principle, be resolved by introducing additional ``hidden'' variables that complete the description and eliminate the indeterminacy.
 Still, the device-independent violation of  Bell inequalities, which is by now a well-established empirical fact \cite{bell1, bell2, bell3, bell4, bell5}, considerably strengthens the idea that quantum mechanics is fundamentally indeterministic. It is, in fact, a mathematical theorem that under very mild assumptions --- i.e., the existence of independent systems (with an arbitrarily small amount of independence) --- any violation of a Bell inequality without signaling implies that the measurement results are random (see e.g. \cite{pironio2010random, acin2016certified, colbeck2012free, putz2014arbitrarily}.) This is an extraordinarily strong result well known to the quantum information science community. 
It appears there are only three plausible ways to deny the existence in physics of random events: 
%It appears the only way to deny the existence in physics of random events is to either: 
%
\begin{itemize}
	\item accept superdeterminism---that is, embrace the radical thesis that everything, down to the smallest detail, was fixed at the Big Bang; or 
	\item accept some form of signaling as in Bohmian mechanics, though this theory hides the signaling behind 
	fundamentally inaccessible so-called nonlocal hidden variables; or
	\item claim that there are no absolute measurement results but that all results happen in parallel, i.e. replace the universe by a multiverse as in the many-world interpretation.
\end{itemize}
Although these alternative interpretations are logically possible, we believe that there is sufficient empirical evidence from quantum theory (and, incidentally, from classical statistical physics and chaos theory) to adopt as a working hypothesis that physical randomness is a real, irreducible feature of the world. This stands in tension with the theory of relativity, in which determinism is typically regarded as a necessary feature.\footnote{
    For a critical discussion, see~\cite{earman2008pruning, del2021relativity} and the references therein.} 
Here, we consider this conflict as an additional motivation to investigate how mathematics and physics can speak of and reason about ontic indeterminacy~\citep{del2024creative}. 

In contrast to quantum theory, classical mechanics is usually regarded as deterministic, since, in almost all cases (see~\cite{norton2003causation}), the dynamical equations of classical mechanics have a unique solution for any given initial conditions. This is correct, but only under the implicit assumption that the initial conditions are fully determined to an infinite precision, i.e., they are represented as a mathematical point in phase space, in turn identified by an $n$-tuple of real numbers. In \cite{born2012physics, reichenbach2012philosophy, del2019physics}, the authors notice that this assumption is not as innocent as one may think at first sight. Indeed, the description of typical real numbers, those not computable by an algorithm, requires an infinite amount of information. If one denies the existence of infinite information --- or merely denies infinite information density in space --- then classical mechanics loses its determinism and manifests indeterminism despite keeping the same dynamical equations. This is especially clear when considering chaotic classical dynamical systems, where infinitesimal digits of the initial condition, initially of no importance at all, quickly become essential for the description of the state of the system. Here, determinism requires that all digits of the initial condition have an ontologically determined value, while assuming finite information implies that not all digits can have a determined value (at least for typical real-valued initial conditions), hence denies determinism. 

Thus, whether considering quantum theory or classical mechanics with finite information, there seems to be a need to develop a mathematical framework that allows one to deal with true (ontological) randomness. This might also be essential to bridge the seemingly incompatible worldviews of relativity and quantum theory, respectively, and eventually help achieve the greater goal of a quantum theory of gravity. 

%As classical mathematics may fail to provide such a framework, it was proposed in \cite{del2019physics} the concept of Finite Information Quantities (FIQs). These mimic real numbers, but, at any finite time, only finitely many digits are fully determined, whereas some further digits are only determined by their chance to eventually gain a specific determined value, and all other digits are assumed fully indeterminate. 
Given that our well-established mathematical theories arguably fail to provide such a framework, the concept of \emph{finite-information quantity}~(FIQ) was originally proposed in \cite{del2019physics} to fill this gap, though still based on classical mathematics. This mimics the concept of real number, but, at any finite time, only finitely many digits are fully determined, whereas some further digits are only determined by their chance to eventually gain a specific determined value, and all other digits are assumed fully indeterminate. 
To achieve that, the authors introduced the concept of \emph{propensity}, which expresses objective, intrinsic tendencies for some events to occur~\citep{del2023prop}. Let $q_j\in [0,1] \cap \Q$ monotonically quantify the tendency of the $j$-th bit of a string to eventually take the value 1. Hence, $q_j=1$ (resp. $0$) means that the $j$-th bit will take value $1$ (resp. $0$) with absolute certainty. If instead a bit has an associated propensity of $1/2$, this means that the bit value is totally indeterminate. 
To be more precise, we call a bit random if its values have not yet been determined. Thus, a bit can be random (indeterminate) or be determined with value $0$ or $1$.

%At some point we need to be more precise: a bit can be
%random (not determined) or be determined with value 0
%or with value 1.

This allows one to formally relax the principle of infinite precision for physical quantities: 
a FIQ is an array of propensities $\{ q_1, q_2, \cdots , q_j, \cdots \}$, such that its information content is finite, i.e.,  $\sum_j I_j < \infty$, where $I_j=1-H(q_j)$ is the information content of the propensity, and $H$ is the binary entropy function of its argument. 

We postulate that physical quantities take values in FIQs instead of in the real numbers. As an example of a physical quantity $\gamma$ that fulfills the above definition of FIQs, take:
\begin{equation}
	\label{fiqs}
	\gamma \left(N(t), M(t)\right)=0.\underbrace{\gamma_1\gamma_2\cdots \gamma_{N(t)}}_{\in \{0, 1\}} \overbrace{q_{N(t)+1}\cdots q_{M(t)}}^{\in(0, 1)\cap \mathbb{Q}}\frac{1}{2}\cdots\frac{1}{2}\cdots.
\end{equation}
At a given time $t$, the first $N$ bits are fully determined (i.e. the propensities are all either $0$ or $1$). The following bits are still indeterminate. Among those, in this simple example, all bits between the $N+1$-th  and  the $M$-th, have a biased tendency to actualize their value to either $0$ or $1$, i.e., the propensities associated to each bits are $q_k\in(0, 1)\cap \mathbb{Q}$. After the $M$-th bit, all the associated propensities are completely random, i.e., $q_{M+1}, q_{M+2}, \dots$ all equal to $1/2$, ensuring finite information content (the information content $I(1/2)=0$).

%This is an alternative to classical mathematics first proposed by Brouwer~\cite{brouwer1907dissertation,brouwer1912intuitionism} featuring a built-in form of indeterminism manifested in the failure of the law of excluded middle: for every proposition $A$, either $A$ or its negation holds.
% \footnote{
%     Constructive mathematics is an alternative to classical mathematics based on the failure of the law of excluded middle: for every proposition $A$, either $A$ or its negation holds. We refer readers to \cite{troelstravandalen1988constructivism,bridges1995constructive} for an introduction.} 

This first attempt to develop a mathematical theory suitable to speak of indeterminism led Gisin~\cite{gisin2020mathematical} directly to intuitionism thanks to Carl Posy. Roughly, constructive mathematics is an umbrella term for any alternative to classical mathematics based on the failure of the law of excluded middle, which states that for every proposition $A$, either $A$ or its negation holds. Intuitionism is a form of constructive mathematics first proposed by Brouwer~\cite{brouwer1907dissertation,brouwer1912intuitionism} featuring built-in randomness manifested in its distinctive treatment of the continuum. 
Still, despite the clear mathematical appeal of intuitionism, it became clear that Brouwer's own justification of it has fundamental limitations: as physicists we look for a realist account of randomness as a ``fact of the world'', while, as it will be explained in~\Cref{sec:brouwer}, Brouwer holds a strong idealist position about mathematics and the exterior world of the subject and attributes indeterminacy to the faculties of a transcendental subject. 
This observation motivated the articulatation of a physicist account of intuitionism~\citep{gisin2021indeterminism}. 
In particular, to accommodate physics, Brouwer's ideas need to be replaced by a power of nature to produce new information in the form of bits that acquire determined values as time processes. Indeed, assuming physical randomness, it is natural to believe that nature possesses such creative power. Once this is accepted, it is fascinating that such a naturalistic understanding of intuitionism provides a solid mathematical framework that allows one to perform most --- possibly all --- of the standard mathematics needed in physics, while simultaneously incorporating at its core a non-geometric view of time sorely missed in classical mathematics. 
We will explore this point in~\Cref{sec:divergences} and illustrate how FIQs can be accommodated intuitionistically.

% This first attempt to develop a mathematical theory suitable to speak of indeterminism led Gisin to encounter intuitionistic mathematics thanks to Carl Posy and subsequently articulate a physicist account of intuitionism~\citep{gisin2021indeterminism}. However, as it will be explained in~\Cref{sec:divergences}, at the same time 
% it became quite clear that Brouwer's original formulation of  intuitionism also has fundamental limitations: as physicists we look for a realist account of randomness as a ``fact of the world'', while Brouwer holds a strong idealist position about mathematics and the exterior world of the subject. 
% In particular, Brouwer's attributes indeterminacy to the faculties of a transcendental subject. 
% %To put it in Heyting's words, ``introspection is useless in physics''~\citep[p.~90]{heyting1974intuitionistic}. 
% To accommodate physics, this needs to be replaced by a power of nature to produce new information in the form of bits that acquire determined values as time processes. Indeed, assuming physical randomness, it is natural to believe that nature possesses such creative power. Once this is accepted, it is fascinating that such a naturalistic understanding of intuitionism provides a solid mathematical framework that allows one to perform most --- possibly all --- of the standard mathematics needed in physics, while simultaneously incorporating at its core a non-geometric view of time sorely missed in classical mathematics. 
% We will explore this point in~\Cref{sec:divergences} and illustrate how FIQs can be accommodated intuitionistically. 

It is entirely plausible that all physical phenomena can be simulated on a classical computer (Turing machine), assuming as we would that the universe has only a finite amount of information. In theory, this can be seen as a corollary of Church's thesis~\cite{Copeland2012-COPTCT}. Given that such machines operate with finite resources, it is equally reasonable to suggest that the laws of physics might be expressible within the framework of intuitionistic mathematics, which inherently reflects the limitations of finitary computation.

From the naturalistic standpoint, numbers are processes, often never ending processes that leave part of the future open and describe natural phenomena like becoming and the creation of fresh information. It is only ``at the end of time'', so to speak, that all processes terminate and hence all mathematical objects, like typical real numbers, gain determined values. But then, it is no surprise that, from the end-of-time point of view, no evolution any longer takes place and that determinism reins. Accordingly, classical mathematics is naturalistic intuitionism seen from the end of time. At any finite time, naturalistic intuitionism might be the right and powerful tool required for physics. Its ability to describe non-geometric time is as intimate as Newton's concept of derivative is to describe velocities. In contrast to Brouwer's idealism, however, physics needs a naturalistic foundation for intuitionism, based solely on the fact that nature has the creative power of randomness.
In the reminder of the paper, we examine how this naturalization of Brouwer's thought can be achieved.

%----------------------------------------------------------------------------------------------------
\section{Brouwer's intuitionism} \label{sec:brouwer}
%----------------------------------------------------------------------------------------------------

It is time to take a step back and overview the original philosophy of mathematics developed by Brouwer~\cite{brouwer1913intuitionism,brouwer1948consciousness,brouwer1981cambridge} through his two acts of intuitionism, so we can be in position to better appreciate the unique features of the naturalization of his intuitionistic mathematics. 
After our presentation of Brouwer's views, we will study in detail how these two acts must be reformulated to incorporate the naturalistic approach sketched above. 
For other recent presentations of Brouwer's two acts, we refer readers to~\cite{posy2020mathematical} or \cite{bentzen2025iep}.

\subsection{The first act of intuitionism} \label{brouwer}

Brouwer developed his intuitionism out of a revised Kantian philosophy of mathematics that renounces spatial intuition in favor of a stronger form of the temporal intuition based on the movement of time as ``one thing which gives way to another thing''. 
His intuitionistic mathematics does not leave room for a separate reality existing independent of the mind. 
In~a~word, Brouwer is an idealist in ontology and truth value~\citep{shapiro2000thinking}. 
Crucially, Brouwer justifies his views by means of his two acts of intuitionism, the first of which construes mathematics as a non-linguistic creative activity of the mind in intuition and lays down the basic ingredients of the discrete: 

\begin{quote}
	{FIRST ACT OF INTUITIONISM} \;
	\textit{Completely separating mathematics from mathematical language and hence from the phenomena of language described by theoretical logic, recognizing that intuitionistic mathematics is an essentially languageless activity of the mind having its origin in the perception of a move of time. This perception of a move of time may be described as the falling apart of a life moment into two distinct things, one of which gives way to the other, but is retained by memory. If the twoity thus born is divested of all quality, it passes into the empty form of the common substratum of all twoities. And it is this common substratum, this empty form, which is the basic intuition of mathematics. }
	\citep[p.~4, emphasis original]{brouwer1981cambridge}
\end{quote}

This passage may sound obscure, but ideas described therein are actually simple. To begin with, it is worth stressing that time here is not understood as scientific time, the usual notion of time analyzed by the laws of physics and measured by our clocks. 
As an idealist, Brouwer thinks of time as one's inner temporal awareness~\cite[p.~99, fn.~2]{brouwer1907dissertation}. 
Moreover, he regards all mathematical objects as mind-dependent  constructions carried out as a product of the intuition of time. 
Like everything else, Brouwer's mathematical universe is also generated in inner time. 
The first object to populate the domain is the number two, namely, the structure ``one thing and then another thing'' that he refers to as the empty twoity. It is intuited by abstraction on our perception of the movement of time. The details are not important for our purposes, but the empty twoity can be regarded as the abstract form of all pairs of a first sensation immediately followed by a second one in time.\footnote{
	See~\cite{bentzen202xtwoity} for an interpretation of Brouwer's intuition of twoity as well as an account of his construction of the positive integers and other discrete objects.}  
Then the number one is created by projection on the first element of the pair described by the empty twoity. All other positive integers, the first infinite ordinal~$\omega$, and all subsequent countable ordinals are generated by iterating this construction process. Finally, we stress that mathematics is independent of logic and language because it is a product of intuition. In principle, it can be done mentally without any need of language for communication, in strong contrast with the dominant trust in formal systems and formalized mathematics today. Brouwer forcefully denies any foundational role for logic in mathematics, against logicists and formalists. 

\subsection{The second act of intuitionism} \label{brouwer2}

%The above characterizes Brouwer's discrete mathematics. 
Brouwer's theory of the continuum is erected on two basic intuitionistic mathematical objects: choice sequences and species. 
As we already saw in the introduction, choice sequences are 
potentially infinite sequences whose elements may be chosen freely or determined by a law. Species are properties that serve as the intuitionistic counterpart of the classical notion of set. 
The crucial difference is that species are intensionally characterized by their defining properties, but sets are extensionally characterized by their elements.\footnote{
	Thus, for example, the species of even numbers and species of non-odd numbers are not one and the same. } 
Choice sequences and species are both introduced to the domain of objects by Brouwer's second act of intuitionism: 
\begin{quote}
	{SECOND ACT OF INTUITIONISM} \;
	\textit{Admitting two ways of creating new mathematical entities: firstly in the shape of more or less freely proceeding infinite sequences of mathematical entities previously acquired (\emph{so that, for example, infinite decimal fractions having neither exact values, nor any guarantee of ever getting exact values are admitted}); secondly in the shape of mathematical species, i.e. properties supposable for mathematical entities previously acquired, satisfying the condition that if they hold for a certain mathematical entity, they also hold for all mathematical entities which have been defined to be `equal' to it, definitions of equality having to satisfy the conditions of symmetry, reflexivity and transitivity.} \citep[p.~8, emphasis original]{brouwer1981cambridge}
\end{quote}
Brouwer stresses that species and choice sequences are an immediate consequence of the self-unfolding of the basic intuition of time~\cite[p.93]{brouwer1981cambridge}. 
Both are essential for doing intuitionistic mathematics, but in this paper we concentrate mostly on choice sequences due to their prominent role in naturalistic intuitionism. 
Indeed, it will be enough for our investigation to make one important remark about species: membership only applies for entities constructed~prior to the definition of a species, thus to extend a species with objects constructed later on one needs to redefine the same species once again. %\footnote{For more this see~\cite{bentzen2025iep}.} 

Call a choice sequence lawlike if it is completely determined by a law, and lawless if it is subject to no law restrictions.\footnote{
	Not every non-lawlike sequence is lawless, since sequences might be only partially determined by a law. } 
The notion of law is simply accepted as primitive. 
Given~that Brouwer views intuitionism as a mental activity, here we can think of laws as humanely computable processes~\cite[\S1.3]{troelstra1977choice}. 
Arguably the most important example of a non-lawlike sequence admitted by Brouwer is one generated by the will of the creating subject, an idealized mind that constructs objects at successive stages with perfect memory, infallible reasoning, and no resource constraints~\citep{vanatten2018creating}. 
For instance, the construction process may consist of the subject continuously picking binary digits (bits) however they like.  
Even though such a sequence is completely lawless, it is still granted the status of mathematical object. As far as we know, there is no direct textual evidence that Brouwer ever accepted sequences generated by events taking place in the world independently of the subject.\footnote{
	Freudenthal reports that Brouwer considered sequences generated by a series of rolls of a die in his lectures~\cite[fn.~4]{vandalen1999counterexamples}. But Brouwer sees the exterior world of the subject as consisting of mind-dependent events~\cite[pp.~1235--1236]{brouwer1948consciousness}. Thus, such sequences remain fundamentally different from the naturalistic sequences to be introduced in~\Cref{nat-continuum}. }

We emphasize two points about choice sequences. First, as the way a choice sequence is constructed is essential to its determination, 
intuitionism distinguishes intensional and extensional criteria of identity for them. 
Two choice sequences~$\alpha$ and $\beta$ are extensionally identical iff for all $n \in \nat$, $\alpha(n) = \beta(n)$. 
By contrast, $\alpha$ and $\beta$ are intensionally identical if they are constructed in the same way. For example, two lawlike sequences are intensionally distinct but extensionally identical if they are given by different laws. Sequences can even be intensionally lawless but extensionally lawlike. For example, the creating subject might make a series of choices $0, 1, 1, 2, 3, 5, 8, 13, 21, 34, ...$ and, at the end of time, when all the choices were already made, something extensionally identical to the Fibonnacci sequence is then accidentally obtained in the complete absence of laws~\citep[\S1.2.1]{bentzen2025iep}. 
Second, at any instant of time, a choice sequence~$\alpha$ only contains finite information encoded in its initial segment or restrictions on its continuation. Thus, to determine whether $\alpha$ has a certain property we never need more than a finite amount of information about $\alpha$. That is, the property will be shared by any other choice sequence that shares this finite information. 
In the intuitionistic literature this is referred to as a continuity principle. 
Such principles can be formulated in different degrees of strength that are not relevant here. 
What is important is that the acceptance of even the weakest form of continuity leads to intuitionistic theorems that refute classical logic altogether~\cite[pp.~60--67]{dummett1977elements}. 
As we shall see in the next section, this incompatibility means that the presence of any form of continuity principle in naturalistic intuitionism would have deep revisionist consequences for any theorem in physics that makes essential use of classical logic or mathematics.

Finally, truth is not a topic that Brouwer discusses in either of his acts of intuitionism, but his so-called creating subject arguments clearly relate the way truths come into being with the construction of choice sequences~\citep{vanatten2018creating}. It thus seems appropriate to say a few things about Brouwer's conception of truth here. 
The creating subject not just constructs mathematical objects but also experiences mathematical truths in time. 
In fact, a proposition is true just when its truth has been experienced (Brouwer, \citeyear{brouwer1948consciousness}, p.~1243; \citeyear{brouwer1955effect}, p.~114). Brouwer does not have an explicit account of proposition. But, as explained by his student, Heyting~\cite{heyting1956intuitionism}, a proposition is understood in terms of its assertability conditions. 
It is thus common to refer to these experiences that realize truths as proofs. Roughly, the meaning of a complex sentence formed by the intuitionistic logical connectives is explained by laying down what a proof of it is~\citep[ch.1 \S3.1]{troelstravandalen1988constructivism}. 	
For example, in this proof explanation, a proof of $A \land B$ is a proof of $A$ and a proof of $B$.  
But we stress that proofs are regarded as mental constructions given in intuition. In Brouwer's intuitionism, there is no room for a subject-transcendent characterization of meaning and truth. 

%----------------------------------------------------------------------------------------------------
\section{The philosophical basis of naturalistic intuitionism} \label{sec:divergences}
%----------------------------------------------------------------------------------------------------

%Brouwer's original two acts of intuitionism

In this section we formulate a basic philosophical framework for naturalistic intuitionism by means of a minimalist reformulation of Brouwer's two acts. At the same time, we explore the unique naturalistic approach to key intuitionistic topics, such as logic, language, time, ontology, meaning, truth, and applications in science. 
In our opinion, this comparative analysis reveals some of the most deep philosophical consequences of the naturalization of intuitionistic mathematics. 

\subsection{Naturalizing the first act of intuitionism} \label{powerofnature} 

Naturalistic intuitionism is erected on a particular notion of external time that is continually producing new information --- again, as existing in the universe and not personal knowledge of an agent. That is, information is not intended as an epistemic concept of the knowledge available to some agent, but rather as something objectively created anew in the universe. 
%\textcolor{red}{(again, as existing in the universe and not personal knowledge of an agent). Note that here information is not intended as an epistemic concept, namely, the knowledge available to some agent, but rather as something objectively created anew in the universe. }
% This is the much-needed non-geometric concept of time characterized by genuine new information generated by indeterminate processes that was alluded to in~\Cref{sec:nat-int}.  
% We shall refer to it here as ``creating time'', as the term expresses the very act of bringing something to life in time.\footnote{
This is the much-needed non-geometric concept of time characterized by genuine new information generated by indeterminate processes that was alluded to in~\Cref{sec:nat-int}.  
We shall refer to it as ``creating time''.\footnote{
	This was first called ``creative time'' in \cite{del2024creative}. It has been renamed here to emphasize action and to match Brouwer's own translation of his Dutch term \textit{het scheppende subject} into English as `the creating subject'~\cite[p.~1246]{brouwer1948consciousness}.}  
This describes the worldview that information is created as the clock ticks. 
The past has already been determined, but future events are left undecided until enough information gets created in time. In fact, we may also say that creating time itself ticks whenever new information is created. 
In what follows, ``time'' refers to creating time unless explicitly stated otherwise.

% How does new information get created in the present? 
% As outlined in~\Cref{sec:nat-int}, one basic tenet of naturalistic intuitionism is that nature has the power to incessantly brings random bits into existence. 
% This power is manifested by a natural random process that repeatedly generates random bits at discrete instants of time.  
% We say that at every instant $n \in \nat$, a fresh random bit, denoted $r(n)$, is determined. 
% Thus, at any instant, only finitely many bits created thus far 
% $r(1), r(2), r(3), ..., r(n)$ 
% have been determined; however, once creating time elapses at the next instant, a fresh bit $r(n+1)$ is determined as well. 
% The process can be repeated indefinitely since creating time can always elapse again. 

How does new information get created in the present? 
As outlined in~\Cref{sec:nat-int}, one basic tenet of naturalistic intuitionism is that nature has the power to incessantly brings random bits into existence. 
This power is manifested by natural random processes that repeatedly generate random bits at discrete instants of time. Let $r$ denote such a natural random process. 
We say that at every instant $n > 0$, a fresh random bit, denoted $r(n)$, is determined. 
Thus, at any instant, only finitely many bits created thus far 
$r(1), r(2), r(3), ..., r(n)$ 
have been determined; however, once creating time elapses at the next instant, a fresh bit $r(n+1)$ is determined as well. 
The process can be repeated indefinitely since creating time can always elapse again. 
This only assumes the existence of binary digits --- labeled $0$ and $1$ --- as basic mathematical objects, the naturalistic counterpart of Brouwer's unity and twoity.\footnote{
	Brouwer's names for the number $1$ and $2$~\cite[p.~90]{brouwer1981cambridge}. } 
%Or rather, these random bits could be regarded as physical objects, since they are real entities that can be somehow observed and interacted with in the world. 
All other numbers can be generated from bits and the power of nature. By regarding algorithms as instances of laws of nature, the construction of any lawlike object is justified. For instance, the number $2$ and all subsequent numbers can be constructed by means of the successor operation out of the bit~$1$. 

In light of all this, we suggest the following paraphrase of Brouwer's first act of intuitionism as a way to better understand the unique traits of naturalistic intuitionism: 

\begin{quote}
	{FIRST ACT OF NATURALISTIC INTUITIONISM} \;
	\textit{Completely separating physics from metaphysical assumptions of determinism and hence from the phenomena of infinite predetermined information, recognizing that intuitionistic 
	mathematics is fundamentally the result of randomness produced by nature 
	having its origin in the move of creating time. This movement of creating time may be described as the creation of new information as time processes, setting an actual separation between before and after, as old information is retained but new information did not exist prior to its creation. Nature has the power to continually produce randomness in the form of entirely new information at discrete instants of creating time. And it is this power, allowing to produce random bits from~$0$ and $1$, which is the basic~concept of naturalistic intuitionism. }
\end{quote}

To sum up, whereas Brouwer's intuitionism starts from the rejection of language and logic as its point of departure, naturalistic intuitionism departs from the rejection of a deterministic worldview grounded on infinite information. 
To borrow Carl Posy's fitting analogy, no infinite helicopter that can lift us high enough to survey the entire terrain or reveal how things are determined at the end of time~\cite[p.~14]{posy2020mathematical}. 
Rather, everything is always evolving in a unfinished process; only what can be computed is completely determined and only the finite is graspable~\citep{gisin2021indeterminism}. On another note, 
Brouwer's intuitionism focuses on inner time while naturalistic intuitionism is interested in outer time. Instead of a basic intuition of mathematics, we have the power of nature as the basic concept for physics. 
This revised first act has a number of significant implications worth exploring: %regarding especially logic and language, temporality, objectivity, and ontology: 

\paragraph{Logic and language} Although it would seem that naturalistic intuitionism is in principle neutral about the role of logic and language in mathematics, the naturalization curiously entails a Brouwerian downplay of these themes as well. It is hard to argue that the power of nature to produce randomness can be fully expressed in words. Mathematics remains a languageless activity, but of nature rather than the mind. Is logic a part of mathematics from the naturalistic viewpoint too? Not necessarily. Naturalistic intuitionism is consistent with, though does not necessarily imply, the view argued by \cite{putnam1969logic} that logic is a natural science. But naturalistically it is also plausible that mathematics depends on logic, in which case nature has to conform to logical laws prescribed in advance. 

%It is the other way around: nature prescribes the rules of the game. 
%Thus, as argued by \cite{putnam1969logic}, logic, insofar as it is a part of mathematics, is a natural science. Given a constructive analytic-synthetic distinction~\citep{martinlof1994analytic,bentzen202xanalytic}, the tautologies of logic can even be said to be synthetic a priori, for they require existence of a proof. We shall see in \Cref{nat-continuum} that proofs can be connected to experiments.

\paragraph{Temporality} Time is the key concept in Brouwer's and naturalistic intuitionism, but it is construed differently in both approaches. For Brouwer, mathematics is a mental creation based on the creating subject's perception of the movement of their internal time, separating between their perception of present and immediate past. In naturalistic intuitionism, mathematics is a product of nature grounded in the movement of creating time. As a model of external time, creating time ticks independently of one's experiences. 
Creating time assumes realism about the past, present and future. 
The future exists as a horizon of real potentialities, which are actualized in the present act of creation characteristic of creating time. 
The next bit to be generated by the power of nature is not yet determined, but 
from the naturalist viewpoint, indeterminate objects have real existence as potentialities. 
This introduces a subtle metaphysical distinction between the real existence of potentialities and well-determinedness of actualities~\cite{del2024creative}. 

\paragraph{Objectivity and communication} We have seen that the naturalization replaces Brouwer's creating subject with the power of nature. This, at the same time, explains the apparent objectivity of mathematics: mathematics deals with real existing objects produced by nature. This means that the common charges of solipsism raised against Brouwer do not affect the naturalistic intuitionist because of the desubjectivization.\footnote{
    It is worth noting that \cite{placek1999mathematical} and \cite[ch.~6]{vanatten2004brouwer} maintain that Brouwerian intuitionistic mathematics has intersubjective validity.} 
One might think the communication of mathematics is not especially problematic in the naturalist setting because mathematicians all speak about the same mind-independent objects. However, lawless sequences are known to resist any linguistic treatment~\cite[p.~5]{vandalen1999role}. 
Thus, after the introduction of choice sequences in the second act, to be revisited in \Cref{nat-continuum}, some objects do become non-communicable despite being mind-independent, simply because they are not yet determined. 

%\textcolor{red}{ELABORATE later}

\paragraph{Ontology} The naturalization induces a novel approach to ontology in intuitionism. 
It is consistent with the Quine--Putnam indispensability argument~\cite[\S1]{sep-mathphil-indis}, for example, unlike Brouwer's intuitionistic mathematics. Naturalistically, mathematical entities are indeed indispensable to our best physical theories and there is no ontological commitment to other dispensable entities such as mental constructions.
The way objects are constructed in both settings is also worth contrasting. 
The basic objects of Brouwer's universe are the number two and one, which, as explained in~\Cref{brouwer}, are constructed from the empty twoity in this very order. 
%There is no analogous of Brouwer's temporal precedence of two over one in the naturalistic approach. The basic objects are labeled $0$ and $1$, which are assumed to have independent existence from one another. In fact, they are not even supposed to represent the movement of time as ``before'' and ``after'', as in Brouwer's intuitionism. If anything, here, as $0$ and $1$ have already been determined in creating time, they are both ``after''.  The only distinction in the movement of time acknowledged in intuitionistic naturalism is that between what has already been determined and what is still to be determined. 
In the naturalistic approach to intuitionism, the basic objects $0$ and $1$ 
%—labeled $0$ and $1$— % 
represent the fundamental distinction between the indeterminate and the determinate; that is, they stand for the ``before'' and the ``after'' an actualization. It is the creative process of actualizing something that was previously only potential which gives rise to the twoity—i.e., in this case, the bit—much in the spirit of Brouwer's temporal perspective, though here without invoking a creating mind.\footnote{
    One can identify another kind of twoity in naturalistic intuitionism. For an event to be fundamentally indeterminate, there must be at least two distinct potential future states, only one of which will actualize.} 

Moreover, the construction of $2$ and all subsequent natural numbers in terms of addition and $1$ is explicitly rejected by Brouwer, while it is accepted naturalistically. As we can see in the quotation below, Brouwer objects that adding two things together already presupposes the intuition of twoity --- such an objection, however, does not apply to the naturalistic intuitionist because addition and any other algorithmic operations on bits are readily available as laws of nature: 

\begin{quote}
	The first act of construction has \textit{two} discrete things thought together [...]
	F. Meyer [...] says that \textit{one} thing is sufficient, because the circumstance that
	I think of it can be added as a second thing; this is false, for exactly
	this \textit{adding} (i.e. setting it while the former is retained) \textit{presupposes
		the intuition of two-ity}; only afterwards this simplest mathematical
	system is projected on the first thing and the ego \textit{which thinks the
		thing}.~\cite[p.~179, fn.~1, emphasis original]{brouwer1907dissertation}
\end{quote}

\paragraph{Science} 

Brouwer originally grounded his intuitionistic views in a mysticist background philosophy that 
describes the external world of the subject 
in terms of iterative complexes of sensations completely estranged from them~\citep[pp.~1235--1236]{brouwer1948consciousness}.\footnote{
		We thank Mark van Atten for bringing to our attention that, at the same time, Brouwer isolates pure mathematics from mysticism, and, as a result, all of science is detached from the realm of mysticism, because Brouwer views science as applied mathematics.} 
Since mathematics belongs to the internal world of the subject, 
it follows that any mathematical techniques needed for physics has to conform to our mind and not nature. In this sense, scientific applications are secondary to Brouwer's program. 
Naturalistic intuitionism, on the other hand, places nature and the needs of science first. 

\subsection{Naturalizing the second act of intuitionism} \label{nat-continuum} 

% Naturalistic intuitionism accepts both species and choice sequences. Due to the naturalization of the first act, however, these objects derive from the power of nature. Indeed, as already mentioned in the previous subsection, the notion of a sequence driven by randomness is absolutely crucial to the naturalistic intuitionist's indeterminsitic physical worldview. Thus, just as in Brouwer's intuitionism, the second act is a consequence of the first. 

Due to the naturalization of the first act, species and choice sequences must derive from the power of nature. Indeed, as mentioned in the previous subsection, the notion of a sequence driven by randomness is absolutely crucial to the naturalistic intuitionist's indeterminsitic physical worldview. Thus, just as in Brouwer's intuitionism, the second act is a consequence of the first. Let us now take a closer look at how the naturalization affects species and choice sequences in turn.

Naturalistic species have subtle differences, given that properties and equivalence relations posited by the laws of nature or natural random processes exist. Thus, they exist independently of any thinking mind, while for Brouwer to be a species is to be intuited. Due to the naturalistic commitment to ontological realism, here a species never has to be redefined to extend its scope to objects constructed after its definition. Once again, determinism and realism must be sharply distinguished. This is not to say that membership in a naturalistic species is decided in advance for objects yet to be constructed. 
As usual, we write $a \in A$ to mean that $a$ has the property described by the naturalistic species $A$. 
Here if $A$ is determined at stage $n$ in time, it is not at this point determined whether $a \in A$, for some new object~$a$ that is determined at stage $k > n$. 
It is only once $a$ is determined at stage~$k$ that $a \in A$ can be determined as well. However, the contrast with Brouwer's intuitionistic mathematics is that $a \in A$ can be determined at $k$ without the need to redefine~$A$ at a further stage~$l > k$ so it can collect $a$ as a previously constructed object. 
In this sense, ontological realism brings about a simplified account of species.
More importantly, every naturalistic species gives rise to natural random processes ranging over them. Thus, if $A$ is a species, we admit all natural random process $r$ such that $r(i) \in A$ for every positive integer $i$. 
So far, we only assumed that each $r(i)$ must have bits as values because the most primitive naturalistic species is that of the bits~$0$ and $1$ provided by nature.

The naturalistic understanding of choice sequences is also remarkably distinct. Here all basic~sequences derive from the power of nature to constantly produce new information in the form of bits whose values are random. 
First of all, we may say every natural random process~$r$ is a lawless generators of bits.\footnote{
	For more on lawless generators and projections see~\cite[p.~68]{posy2020mathematical}. } 
Recall that at any discrete instant of time~$n \in \nat$, a fresh random bit, denoted $r(n)$, is determined. 
Naturalistically, the most basic construction of a choice sequence begins by noting that, at time $n$, we also have at our disposal a finite sequence of previously created bits $r (1), r (2), . . . , r (n)$ that are completely independent from each other. 
If this process is continued indefinitely, without any further intervention, the resulting object is a lawless sequence of bits. 
But more generally, in naturalistic intuitionism, any expressible non-lawlike sequence
$$\alpha=\langle \alpha(1), \alpha(2), ... \rangle$$
of arbitrary mathematical objects is generated in the following way~\cite[\S3]{gisin2021indeterminism}. Given a fixed object~$a$ and a law~$f$ that takes as arguments the sequence of bits generated thus far, the number of such bits, and the previous element of the sequence, we have:\footnote{
	Depending on the law $f$ adopted, choice sequences of computable numbers need not converge. In \cite{gisin2021indeterminism} it was assumed that they all do. }
\begin{align*}
	\alpha(1) &= a \\
	\alpha(n+1) &= f(r(1), ..., r(n), n, \alpha(n)).
\end{align*}
Naturalistically, the most fundamental sequences are all obtained in this way as projections of lawless sequences of bits to mathematical objects according to a law. 

We stress three distinguishing aspects of these so-called ``naturalistic sequences''. 
First, the process described above only admits intensionally lawless sequences if they range over bits --- in fact, the only intensionally lawless sequence is the natural random process itself. 
Recall the distinction made in \Cref{brouwer2} about intensional and extensional identity. 
All lawless sequences of other objects are generated as projections partially determined by laws, so they can only be extensionally and not intensionally lawless. 
% \footnote{In contrast to this,
% 	two sequences formed by the same initial object $a$ and the same laws might be extensionally different (because the random bits differ) but remain in a sense ``intensionally alike''.	
% } 
%%
Second, as in Brouwer's intuitionism, a law is a primitive notion of computation process. However, we stress that here they are not processes constructed by some constructing intelligence, who lets them come into being as they are defined, but generated as instances of laws of nature. The existence of laws is predetermined by nature. 
Third, the FIQs introduced in~\Cref{sec:nat-int} can be accommodated as a specific type of naturalistic sequence, in which the next bit is generated based on a majority rule applied to the preceding $k$ bits, where $k$ is an odd number. Within this framework, the propensities characterizing FIQs emerge from the inherent randomness of making binary choices with equal probability (i.e., 50\%--50\%). When a significant majority of the last $k$ bits are $1$s, there is an increased chance that this pattern will continue.

To incorporate the unique treatment of species and choice sequences described above, it seems reasonable to reformulate Brouwer's second act as follows: 

\begin{quote}
{SECOND ACT OF NATURALISTIC INTUITIONISM} \;
\textit{Admitting two ways of creating new mathematical entities: firstly in the shape of sequences generated from natural random processes according to the laws of nature (\emph{so that, for example, FIQs having neither determinate values, nor any guarantee of ever getting determinate values are admitted}); secondly in the shape of mathematical species, i.e. properties preserving some equivalence relation that define membership equality in the species, where the property and equivalence relation must be describable by the laws of nature or its random processes.}
\end{quote}

\paragraph{Determinism and realism} 

Non-lawlike naturalistic sequences are not as determined as lawlike sequences. But the potentialities of each subsequent choice are real, hence here non-lawlike sequences are real objects in their entirety, but not yet fully determined. We stress that naturalistic intuitionism sharply distinguishes between determinism and realism: the later does not imply the former. 
Objects may have real existence but possess only partial determinacy at finite time.

\paragraph{Choice}

A natural question is whether the same class of choice sequences is expressible in both naturalistic and Brouwer's intuitionism. 
It is easy to see that extensionally the same class of choice sequences is obtained. 
For every naturalist choice sequence~$\alpha=\langle \alpha(1), \alpha(2), ... \rangle$, the creating subject may construct an extensionally identical choice sequence in Brouwer's intuitionism by choosing its elements step by step. 
Conversely, given any choice sequence~$\beta=\langle \beta(1), \beta(2), ... \rangle$ in Brouwer's intuitionism, there is a natural random process $r$ ranging over the species collecting the elements of the sequence that is extensionally equal to it. 
% \textcolor{red}{Conversely, given any choice sequence~$\beta=\langle \beta(1), \beta(2), ... \rangle$ in Brouwer's intuitionism, there is a sequence in naturalistic intuitionism constructed by setting $\alpha(1) = \beta(1)$ and for every $\alpha(n+1)$ we adopt as our $f$ the law given by $r(1), ..., r(n), n, \alpha(n) \mapsto \beta(n+1)$. }
%%
But these sequences need not be intensionally identical because the construction processes may differ. There are no choice sequences generated by the creating subject in the naturalist account.

\paragraph{Constructivity}

One might wonder if to do physics one needs the full power of Brouwer's intuitionistic mathematics, which as already mentioned in \Cref{brouwer2} includes classically false additional principles to reason about choice sequences. Would the form of constructive mathematics practiced by~\cite{bishop1967foundations}, which is roughly equivalent to classical mathematics without the law of excluded middle, not already result in the kind of indeterminacy one needs for physics? Bishop only admits lawlike sequences. 
From the naturalist perspective studied here, only choice sequences allow mathematics to truly speak about randomness because not all sequences found in nature are lawlike. 
At the same time, this raises the question of what exact additional principles that naturalistic choice sequences are expected to satisfy. Two things can be said on this regard. 
First, there is no reason to assume that the naturalistic approach needs to accept all principles employed by Brouwer and his followers. Second, naturalistic intuitionism refrains from settling the question once for all and endows their choice sequences with an open-ended nature. Any additional principles are acceptable to the extent that they found to be useful to do physics or any other science. 

\paragraph{Meaning and truth} 

%Dummett's~\cite{dummett1975philosophical}

In Brouwer, meaning and truth are subject-dependent notions. 
The naturalist rejection of his creating subject implies a complete desubjectification similar in effect to the kind advocated by~Dummett's~\cite{dummett1975philosophical} meaning-theoretical turn: meaning and truth are explained by appealing to a primitive notion of provability no longer reducible to a mental construction in intuition. 
However, naturalistically some proofs can be interpreted as empirical evidence. More specifically, the meaning of a sentence in the naturalist setting is still determined by the proof explanation as in~\Cref{brouwer2}. 
But some proofs are observable information that verify a given proposition. 
We may therefore think of a proof as comprising a physics experiment broadly understood~\cite[p.~5]{martin2014truth}. 
For example, an experiment for ``the temperature now in Geneva is $25^{\circ} \text{C}$'' could be a thermometer showing the designated temperature at the relevant time and place. Mathematical sentences are treated similarly: an experiment for ``$5+7 = 12$'' may be simply counting the relevant units.

How should truth be understood in naturalistic intuitionism? We are looking for some subject-independent account in terms of provability. %Several options have already been explored in the literature in recent years. 
Perhaps following a distinction extensively studied by Raatikainen~\cite{raatikainen2004conceptions}, one may be tempted to frame the discussion through a potentialist-actualist distinction that identifies ``$A$ is true'' with either ``$A$ can be proved'' or ``$A$ has been proved''. 
Or one might wish to further distinguish intermediate conceptions of truth like Hanson~\cite{hansen2016brouwer} does. 
However, recall that naturalistic intuitionism maintains a worldview where things are evolving as the clock ticks. 
Time and fresh information, in the form of new bits, are intimately connected. They go hand in hand: time ticks when new bits are created. Accordingly, as long as time passes, new information is created and new evidence becomes available in the world. 
As empirical evidence, proofs are real objects interacting with the world and are thus processes developing in time. 	
As a result, there is no room for an atemporal conception of truth. We thus have: 

\begin{center}
$A$ is true now = $A$ can be proved now (using information available at present)
\end{center}

If Goldbach's conjecture is eventually proved tomorrow, for example, was it already true prior to that moment according to this naturalistic rearticulation of intuitionistic truth? 
More generally, we stress that if $A$ is true at time $n$, it will remain true at $n+1$. But it need not be the case that $A$ was already true at stage $n-1$, for example. The only exception is when we are dealing with a decidable proposition. In this case, we can at stage $n-1$ prove $A$ using information available at present even if the proof is only given at stage $n$. 
Finally, we add that this account of truth is consistent with realism about proofs. The proof~$a$ as an object in development itself existed prior to that stage $n$ in time, but the truth of the proposition $A$ requires that $a$ proves $A$. This could not have been determined in advance because the proof~$a$ \textit{qua} experiment was still under construction. 

%----------------------------------------------------------------------------------------------------
\section{Concluding remarks} \label{sec:conclusion}
%----------------------------------------------------------------------------------------------------

If nature can be simulated by a classical computer (as suggested by Church's thesis), then a form of mathematics that unfolds over time while remaining finite at each moment may offer a more suitable framework for describing physical phenomena. Motivated by this idea, we have presented naturalistic intuitionism as a novel program in the philosophy of mathematics, grounded in the foundational needs of physics. More importantly, we have pinpointed the necessary revisions to the two acts that uphold Brouwer's intuitionism in order to establish the philosophical foundation backing this program, and have delved into some of the implications arising from this.

%More importantly, we have identified what revisions to the two acts that support Brouwer's intuitionism are necessary to obtain the philosophical background supporting this program and explored some of these implications. 

Several important questions still remain open. In particular, just how much of the classical mathematical results ordinarily needed for doing physics are we able to retain assuming the naturalist intuitionisitic revision discussed in this paper? Since intuitionistic and classical mathematics are inconsistent with each other, the matter should be treated carefully. Also, from naturalistic viewpoint presented in~\ref{sec:nat-int}, classical mathematics is like intuitionism at ``infinite time'' or looked at form the ``end of time''. Since they are inconsistent, it means that there is a singular limit, where truth values suddenly get determined at infinity. 

\section*{Acknowledgments}
We thank Christian W\"uthrich and Mark van Atten for useful comments. F.D.S. acknowledges support from  FWF (Austrian Science Fund) through an Erwin Schrödinger Fellowship (Project J 4699). NG acknowledges support by he Swiss NCCR SwissMap.

\bibliography{biblio}

%apsrev4-2.bst 2019-01-14 (MD) hand-edited version of apsrev4-1.bst
%Control: key (0)
%Control: author (8) initials jnrlst
%Control: editor formatted (1) identically to author
%Control: production of article title (0) allowed
%Control: page (0) single
%Control: year (1) truncated
%Control: production of eprint (0) enabled
\begin{thebibliography}{64}%
\makeatletter
\providecommand \@ifxundefined [1]{%
 \@ifx{#1\undefined}
}%
\providecommand \@ifnum [1]{%
 \ifnum #1\expandafter \@firstoftwo
 \else \expandafter \@secondoftwo
 \fi
}%
\providecommand \@ifx [1]{%
 \ifx #1\expandafter \@firstoftwo
 \else \expandafter \@secondoftwo
 \fi
}%
\providecommand \natexlab [1]{#1}%
\providecommand \enquote  [1]{``#1''}%
\providecommand \bibnamefont  [1]{#1}%
\providecommand \bibfnamefont [1]{#1}%
\providecommand \citenamefont [1]{#1}%
\providecommand \href@noop [0]{\@secondoftwo}%
\providecommand \href [0]{\begingroup \@sanitize@url \@href}%
\providecommand \@href[1]{\@@startlink{#1}\@@href}%
\providecommand \@@href[1]{\endgroup#1\@@endlink}%
\providecommand \@sanitize@url [0]{\catcode `\\12\catcode `\$12\catcode `\&12\catcode `\#12\catcode `\^12\catcode `\_12\catcode `\%12\relax}%
\providecommand \@@startlink[1]{}%
\providecommand \@@endlink[0]{}%
\providecommand \url  [0]{\begingroup\@sanitize@url \@url }%
\providecommand \@url [1]{\endgroup\@href {#1}{\urlprefix }}%
\providecommand \urlprefix  [0]{URL }%
\providecommand \Eprint [0]{\href }%
\providecommand \doibase [0]{https://doi.org/}%
\providecommand \selectlanguage [0]{\@gobble}%
\providecommand \bibinfo  [0]{\@secondoftwo}%
\providecommand \bibfield  [0]{\@secondoftwo}%
\providecommand \translation [1]{[#1]}%
\providecommand \BibitemOpen [0]{}%
\providecommand \bibitemStop [0]{}%
\providecommand \bibitemNoStop [0]{.\EOS\space}%
\providecommand \EOS [0]{\spacefactor3000\relax}%
\providecommand \BibitemShut  [1]{\csname bibitem#1\endcsname}%
\let\auto@bib@innerbib\@empty
%</preamble>
\bibitem [{\citenamefont {Hellman}(1993{\natexlab{a}})}]{hellman1993gleason}%
  \BibitemOpen
  \bibfield  {author} {\bibinfo {author} {\bibfnamefont {G.}~\bibnamefont {Hellman}},\ }\bibfield  {title} {\bibinfo {title} {Gleason's theorem is not constructively provable},\ }\href {https://doi.org/10.1007/BF01049261} {\bibfield  {journal} {\bibinfo  {journal} {Journal of Philosophical Logic}\ ,\ \bibinfo {pages} {193}} (\bibinfo {year} {1993}{\natexlab{a}})}\BibitemShut {NoStop}%
\bibitem [{\citenamefont {Hellman}(1993{\natexlab{b}})}]{hellman1993constructive}%
  \BibitemOpen
  \bibfield  {author} {\bibinfo {author} {\bibfnamefont {G.}~\bibnamefont {Hellman}},\ }\bibfield  {title} {\bibinfo {title} {Constructive mathematics and quantum mechanics: Unbounded operators and the spectral theorem},\ }\href {https://doi.org/10.1007/BF01049303} {\bibfield  {journal} {\bibinfo  {journal} {Journal of Philosophical Logic}\ ,\ \bibinfo {pages} {221}} (\bibinfo {year} {1993}{\natexlab{b}})}\BibitemShut {NoStop}%
\bibitem [{\citenamefont {Bridges}(1995)}]{bridges1995constructive}%
  \BibitemOpen
  \bibfield  {author} {\bibinfo {author} {\bibfnamefont {D.~S.}\ \bibnamefont {Bridges}},\ }\bibfield  {title} {\bibinfo {title} {Constructive mathematics and unbounded operators—a reply to hellman},\ }\href {https://doi.org/10.1007/BF01052602} {\bibfield  {journal} {\bibinfo  {journal} {Journal of Philosophical Logic}\ }\textbf {\bibinfo {volume} {24}},\ \bibinfo {pages} {549} (\bibinfo {year} {1995})}\BibitemShut {NoStop}%
\bibitem [{\citenamefont {Billinge}(1997)}]{billinge1997constructive}%
  \BibitemOpen
  \bibfield  {author} {\bibinfo {author} {\bibfnamefont {H.}~\bibnamefont {Billinge}},\ }\bibfield  {title} {\bibinfo {title} {A constructive formulation of {Gleason}'s theorem},\ }\href {https://doi.org/10.1023/A:1004275113665} {\bibfield  {journal} {\bibinfo  {journal} {Journal of philosophical logic}\ }\textbf {\bibinfo {volume} {26}},\ \bibinfo {pages} {661} (\bibinfo {year} {1997})}\BibitemShut {NoStop}%
\bibitem [{\citenamefont {Richman}(2000)}]{richman2000gleason}%
  \BibitemOpen
  \bibfield  {author} {\bibinfo {author} {\bibfnamefont {F.}~\bibnamefont {Richman}},\ }\bibfield  {title} {\bibinfo {title} {Gleason's theorem has a constructive proof},\ }\href {https://doi.org/10.1023/A:1004791723301} {\bibfield  {journal} {\bibinfo  {journal} {Journal of Philosophical Logic}\ }\textbf {\bibinfo {volume} {29}},\ \bibinfo {pages} {425} (\bibinfo {year} {2000})}\BibitemShut {NoStop}%
\bibitem [{\citenamefont {Fletcher}(2002)}]{fletcher2002constructivist}%
  \BibitemOpen
  \bibfield  {author} {\bibinfo {author} {\bibfnamefont {P.}~\bibnamefont {Fletcher}},\ }\bibfield  {title} {\bibinfo {title} {A constructivist perspective on physics},\ }\href {https://doi.org/10.1093/philmat/10.1.26} {\bibfield  {journal} {\bibinfo  {journal} {Philosophia Mathematica}\ }\textbf {\bibinfo {volume} {10}},\ \bibinfo {pages} {26} (\bibinfo {year} {2002})}\BibitemShut {NoStop}%
\bibitem [{\citenamefont {Gisin}(2021{\natexlab{a}})}]{gisin2021indeterminism}%
  \BibitemOpen
  \bibfield  {author} {\bibinfo {author} {\bibfnamefont {N.}~\bibnamefont {Gisin}},\ }\bibfield  {title} {\bibinfo {title} {Indeterminism in physics and intuitionistic mathematics},\ }\href@noop {} {\bibfield  {journal} {\bibinfo  {journal} {Synthese}\ }\textbf {\bibinfo {volume} {199}},\ \bibinfo {pages} {13345} (\bibinfo {year} {2021}{\natexlab{a}})}\BibitemShut {NoStop}%
\bibitem [{\citenamefont {Del~Santo}(2018)}]{del2018striving}%
  \BibitemOpen
  \bibfield  {author} {\bibinfo {author} {\bibfnamefont {F.}~\bibnamefont {Del~Santo}},\ }\bibfield  {title} {\bibinfo {title} {Striving for realism, not for determinism: Historical misconceptions on {E}instein and {B}ohm},\ }\href@noop {} {\bibfield  {journal} {\bibinfo  {journal} {APS News}\ } (\bibinfo {year} {2018})}\BibitemShut {NoStop}%
\bibitem [{\citenamefont {Del~Santo}\ and\ \citenamefont {Gisin}(2019)}]{del2019physics}%
  \BibitemOpen
  \bibfield  {author} {\bibinfo {author} {\bibfnamefont {F.}~\bibnamefont {Del~Santo}}\ and\ \bibinfo {author} {\bibfnamefont {N.}~\bibnamefont {Gisin}},\ }\bibfield  {title} {\bibinfo {title} {Physics without determinism: Alternative interpretations of classical physics},\ }\href@noop {} {\bibfield  {journal} {\bibinfo  {journal} {Physical Review A}\ }\textbf {\bibinfo {volume} {100}},\ \bibinfo {pages} {062107} (\bibinfo {year} {2019})}\BibitemShut {NoStop}%
\bibitem [{\citenamefont {Gisin}(2020{\natexlab{a}})}]{gisin2020real}%
  \BibitemOpen
  \bibfield  {author} {\bibinfo {author} {\bibfnamefont {N.}~\bibnamefont {Gisin}},\ }\bibfield  {title} {\bibinfo {title} {Real numbers are the hidden variables of classical mechanics},\ }\href@noop {} {\bibfield  {journal} {\bibinfo  {journal} {Quantum Studies: Mathematics and Foundations}\ }\textbf {\bibinfo {volume} {7}},\ \bibinfo {pages} {197} (\bibinfo {year} {2020}{\natexlab{a}})}\BibitemShut {NoStop}%
\bibitem [{\citenamefont {Gisin}(2021{\natexlab{b}})}]{gisin2021}%
  \BibitemOpen
  \bibfield  {author} {\bibinfo {author} {\bibfnamefont {N.}~\bibnamefont {Gisin}},\ }\bibfield  {title} {\bibinfo {title} {Indeterminism in physics, classical chaos and {B}ohmian mechanics: Are real numbers really real?},\ }\href {https://doi.org/https://doi.org/10.1007/s10670-019-00165-8} {\bibfield  {journal} {\bibinfo  {journal} {Erkenntnis}\ }\textbf {\bibinfo {volume} {86}},\ \bibinfo {pages} {1469} (\bibinfo {year} {2021}{\natexlab{b}})}\BibitemShut {NoStop}%
\bibitem [{\citenamefont {Del~Santo}\ and\ \citenamefont {Gisin}(2021)}]{del2021relativity}%
  \BibitemOpen
  \bibfield  {author} {\bibinfo {author} {\bibfnamefont {F.}~\bibnamefont {Del~Santo}}\ and\ \bibinfo {author} {\bibfnamefont {N.}~\bibnamefont {Gisin}},\ }\bibfield  {title} {\bibinfo {title} {The relativity of indeterminacy},\ }\href@noop {} {\bibfield  {journal} {\bibinfo  {journal} {Entropy}\ }\textbf {\bibinfo {volume} {23}},\ \bibinfo {pages} {1326} (\bibinfo {year} {2021})}\BibitemShut {NoStop}%
\bibitem [{\citenamefont {Del~Santo}(2021)}]{del2021indeterminism}%
  \BibitemOpen
  \bibfield  {author} {\bibinfo {author} {\bibfnamefont {F.}~\bibnamefont {Del~Santo}},\ }\bibinfo {title} {Indeterminism, causality and information: Has physics ever been deterministic?},\ in\ \href {https://doi.org/10.1007/978-3-030-70354-7_5} {\emph {\bibinfo {booktitle} {Undecidability, Uncomputability, and Unpredictability}}},\ \bibinfo {editor} {edited by\ \bibinfo {editor} {\bibfnamefont {A.}~\bibnamefont {Aguirre}}, \bibinfo {editor} {\bibfnamefont {Z.}~\bibnamefont {Merali}},\ and\ \bibinfo {editor} {\bibfnamefont {D.}~\bibnamefont {Sloan}}}\ (\bibinfo  {publisher} {Springer International Publishing},\ \bibinfo {address} {Cham},\ \bibinfo {year} {2021})\ pp.\ \bibinfo {pages} {63--79}\BibitemShut {NoStop}%
\bibitem [{\citenamefont {Del~Santo}\ and\ \citenamefont {Gisin}(2023{\natexlab{a}})}]{del2023prop}%
  \BibitemOpen
  \bibfield  {author} {\bibinfo {author} {\bibfnamefont {F.}~\bibnamefont {Del~Santo}}\ and\ \bibinfo {author} {\bibfnamefont {N.}~\bibnamefont {Gisin}},\ }\bibfield  {title} {\bibinfo {title} {Potentiality realism: a realistic and indeterministic physics based on propensities},\ }\href@noop {} {\bibfield  {journal} {\bibinfo  {journal} {Euro Jnl Phil Sci}\ }\textbf {\bibinfo {volume} {13}} (\bibinfo {year} {2023}{\natexlab{a}})}\BibitemShut {NoStop}%
\bibitem [{\citenamefont {Del~Santo}\ and\ \citenamefont {Gisin}(2023{\natexlab{b}})}]{santo2023open}%
  \BibitemOpen
  \bibfield  {author} {\bibinfo {author} {\bibfnamefont {F.}~\bibnamefont {Del~Santo}}\ and\ \bibinfo {author} {\bibfnamefont {N.}~\bibnamefont {Gisin}},\ }\bibfield  {title} {\bibinfo {title} {The open past in an indeterministic physics},\ }\href@noop {} {\bibfield  {journal} {\bibinfo  {journal} {Foundations of Physics}\ }\textbf {\bibinfo {volume} {53}},\ \bibinfo {pages} {4} (\bibinfo {year} {2023}{\natexlab{b}})}\BibitemShut {NoStop}%
\bibitem [{\citenamefont {Del~Santo}\ and\ \citenamefont {Gisin}(2024)}]{del2024creative}%
  \BibitemOpen
  \bibfield  {author} {\bibinfo {author} {\bibfnamefont {F.}~\bibnamefont {Del~Santo}}\ and\ \bibinfo {author} {\bibfnamefont {N.}~\bibnamefont {Gisin}},\ }\bibfield  {title} {\bibinfo {title} {Creative and geometric times in physics, mathematics, logic, and philosophy},\ }\href@noop {} {\bibfield  {journal} {\bibinfo  {journal} {arXiv preprint arXiv:2404.06566}\ } (\bibinfo {year} {2024})}\BibitemShut {NoStop}%
\bibitem [{\citenamefont {Del~Santo}\ and\ \citenamefont {Gisin}(2025)}]{del2024features}%
  \BibitemOpen
  \bibfield  {author} {\bibinfo {author} {\bibfnamefont {F.}~\bibnamefont {Del~Santo}}\ and\ \bibinfo {author} {\bibfnamefont {N.}~\bibnamefont {Gisin}},\ }\bibfield  {title} {\bibinfo {title} {Which features of quantum physics are not fundamentally quantum but are due to indeterminism?},\ }\href@noop {} {\bibfield  {journal} {\bibinfo  {journal} {Quantum}\ }\textbf {\bibinfo {volume} {9}},\ \bibinfo {pages} {1686} (\bibinfo {year} {2025})}\BibitemShut {NoStop}%
\bibitem [{\citenamefont {Eyink}\ and\ \citenamefont {Bandak}(2020)}]{eyink2020renormalization}%
  \BibitemOpen
  \bibfield  {author} {\bibinfo {author} {\bibfnamefont {G.~L.}\ \bibnamefont {Eyink}}\ and\ \bibinfo {author} {\bibfnamefont {D.}~\bibnamefont {Bandak}},\ }\bibfield  {title} {\bibinfo {title} {Renormalization group approach to spontaneous stochasticity},\ }\href@noop {} {\bibfield  {journal} {\bibinfo  {journal} {Physical Review Research}\ }\textbf {\bibinfo {volume} {2}},\ \bibinfo {pages} {043161} (\bibinfo {year} {2020})}\BibitemShut {NoStop}%
\bibitem [{\citenamefont {Ben-Yami}(2020)}]{ben2020structure}%
  \BibitemOpen
  \bibfield  {author} {\bibinfo {author} {\bibfnamefont {H.}~\bibnamefont {Ben-Yami}},\ }\bibfield  {title} {\bibinfo {title} {The structure of space and time, and physical indeterminacy},\ }\href@noop {} {\bibfield  {journal} {\bibinfo  {journal} {arXiv preprint arXiv:2005.05121}\ } (\bibinfo {year} {2020})}\BibitemShut {NoStop}%
\bibitem [{\citenamefont {Dolev}(2024)}]{dolev2024temporal}%
  \BibitemOpen
  \bibfield  {author} {\bibinfo {author} {\bibfnamefont {Y.}~\bibnamefont {Dolev}},\ }\bibfield  {title} {\bibinfo {title} {Temporal direction, intuitionism and physics},\ }\href@noop {} {\bibfield  {journal} {\bibinfo  {journal} {Entropy}\ }\textbf {\bibinfo {volume} {26}},\ \bibinfo {pages} {594} (\bibinfo {year} {2024})}\BibitemShut {NoStop}%
\bibitem [{\citenamefont {Brouwer}(1975)}]{brouwer1975collected}%
  \BibitemOpen
  \bibfield  {author} {\bibinfo {author} {\bibfnamefont {L.~E.~J.}\ \bibnamefont {Brouwer}},\ }\href {https://doi.org/10.1016/C2013-0-11893-4} {\emph {\bibinfo {title} {{L. E. J. Brouwer Collected Works 1. Philosophy and Foundations of Mathematics}}}}\ (\bibinfo  {publisher} {North-Holland},\ \bibinfo {address} {Amsterdam},\ \bibinfo {year} {1975})\ \bibinfo {note} {edited by A. Heyting}\BibitemShut {NoStop}%
\bibitem [{\citenamefont {Brouwer}(1981)}]{brouwer1981cambridge}%
  \BibitemOpen
  \bibfield  {author} {\bibinfo {author} {\bibfnamefont {L.~E.~J.}\ \bibnamefont {Brouwer}},\ }\href@noop {} {\emph {\bibinfo {title} {Brouwer's Cambridge Lectures on Intuitionism}}}\ (\bibinfo  {publisher} {Cambridge University Press},\ \bibinfo {address} {Cambridge},\ \bibinfo {year} {1981})\ \bibinfo {note} {edited by D. van Dalen}\BibitemShut {NoStop}%
\bibitem [{\citenamefont {Heyting}(1974)}]{heyting1974intuitionistic}%
  \BibitemOpen
  \bibfield  {author} {\bibinfo {author} {\bibfnamefont {A.}~\bibnamefont {Heyting}},\ }\bibfield  {title} {\bibinfo {title} {Intuitionistic views on the nature of mathematics},\ }\href@noop {} {\bibfield  {journal} {\bibinfo  {journal} {Synthese}\ }\textbf {\bibinfo {volume} {27}},\ \bibinfo {pages} {79} (\bibinfo {year} {1974})}\BibitemShut {NoStop}%
\bibitem [{\citenamefont {Dolev}(2018)}]{dolev2018physics}%
  \BibitemOpen
  \bibfield  {author} {\bibinfo {author} {\bibfnamefont {Y.}~\bibnamefont {Dolev}},\ }\bibfield  {title} {\bibinfo {title} {Physics’ silence on time},\ }\href@noop {} {\bibfield  {journal} {\bibinfo  {journal} {European Journal for Philosophy of Science}\ }\textbf {\bibinfo {volume} {8}},\ \bibinfo {pages} {455} (\bibinfo {year} {2018})}\BibitemShut {NoStop}%
\bibitem [{\citenamefont {Hensen}\ \emph {et~al.}(2015)\citenamefont {Hensen}, \citenamefont {Bernien}, \citenamefont {Dr{\'e}au}, \citenamefont {Reiserer}, \citenamefont {Kalb}, \citenamefont {Blok}, \citenamefont {Ruitenberg}, \citenamefont {Vermeulen}, \citenamefont {Schouten}, \citenamefont {Abell{\'a}n} \emph {et~al.}}]{bell1}%
  \BibitemOpen
  \bibfield  {author} {\bibinfo {author} {\bibfnamefont {B.}~\bibnamefont {Hensen}}, \bibinfo {author} {\bibfnamefont {H.}~\bibnamefont {Bernien}}, \bibinfo {author} {\bibfnamefont {A.~E.}\ \bibnamefont {Dr{\'e}au}}, \bibinfo {author} {\bibfnamefont {A.}~\bibnamefont {Reiserer}}, \bibinfo {author} {\bibfnamefont {N.}~\bibnamefont {Kalb}}, \bibinfo {author} {\bibfnamefont {M.~S.}\ \bibnamefont {Blok}}, \bibinfo {author} {\bibfnamefont {J.}~\bibnamefont {Ruitenberg}}, \bibinfo {author} {\bibfnamefont {R.~F.}\ \bibnamefont {Vermeulen}}, \bibinfo {author} {\bibfnamefont {R.~N.}\ \bibnamefont {Schouten}}, \bibinfo {author} {\bibfnamefont {C.}~\bibnamefont {Abell{\'a}n}}, \emph {et~al.},\ }\bibfield  {title} {\bibinfo {title} {Loophole-free {Bell} inequality violation using electron spins separated by 1.3 kilometres},\ }\href@noop {} {\bibfield  {journal} {\bibinfo  {journal} {Nature}\ }\textbf {\bibinfo {volume} {526}},\ \bibinfo {pages} {682} (\bibinfo {year} {2015})}\BibitemShut {NoStop}%
\bibitem [{\citenamefont {Shalm}\ \emph {et~al.}(2015)\citenamefont {Shalm}, \citenamefont {Meyer-Scott}, \citenamefont {Christensen}, \citenamefont {Bierhorst}, \citenamefont {Wayne}, \citenamefont {Stevens}, \citenamefont {Gerrits}, \citenamefont {Glancy}, \citenamefont {Hamel}, \citenamefont {Allman} \emph {et~al.}}]{bell2}%
  \BibitemOpen
  \bibfield  {author} {\bibinfo {author} {\bibfnamefont {L.~K.}\ \bibnamefont {Shalm}}, \bibinfo {author} {\bibfnamefont {E.}~\bibnamefont {Meyer-Scott}}, \bibinfo {author} {\bibfnamefont {B.~G.}\ \bibnamefont {Christensen}}, \bibinfo {author} {\bibfnamefont {P.}~\bibnamefont {Bierhorst}}, \bibinfo {author} {\bibfnamefont {M.~A.}\ \bibnamefont {Wayne}}, \bibinfo {author} {\bibfnamefont {M.~J.}\ \bibnamefont {Stevens}}, \bibinfo {author} {\bibfnamefont {T.}~\bibnamefont {Gerrits}}, \bibinfo {author} {\bibfnamefont {S.}~\bibnamefont {Glancy}}, \bibinfo {author} {\bibfnamefont {D.~R.}\ \bibnamefont {Hamel}}, \bibinfo {author} {\bibfnamefont {M.~S.}\ \bibnamefont {Allman}}, \emph {et~al.},\ }\bibfield  {title} {\bibinfo {title} {Strong loophole-free test of local realism},\ }\href@noop {} {\bibfield  {journal} {\bibinfo  {journal} {Physical review letters}\ }\textbf {\bibinfo {volume} {115}},\ \bibinfo {pages} {250402} (\bibinfo {year} {2015})}\BibitemShut {NoStop}%
\bibitem [{\citenamefont {Giustina}\ \emph {et~al.}(2015)\citenamefont {Giustina}, \citenamefont {Versteegh}, \citenamefont {Wengerowsky}, \citenamefont {Handsteiner}, \citenamefont {Hochrainer}, \citenamefont {Phelan}, \citenamefont {Steinlechner}, \citenamefont {Kofler}, \citenamefont {Larsson}, \citenamefont {Abell{\'a}n} \emph {et~al.}}]{bell3}%
  \BibitemOpen
  \bibfield  {author} {\bibinfo {author} {\bibfnamefont {M.}~\bibnamefont {Giustina}}, \bibinfo {author} {\bibfnamefont {M.~A.}\ \bibnamefont {Versteegh}}, \bibinfo {author} {\bibfnamefont {S.}~\bibnamefont {Wengerowsky}}, \bibinfo {author} {\bibfnamefont {J.}~\bibnamefont {Handsteiner}}, \bibinfo {author} {\bibfnamefont {A.}~\bibnamefont {Hochrainer}}, \bibinfo {author} {\bibfnamefont {K.}~\bibnamefont {Phelan}}, \bibinfo {author} {\bibfnamefont {F.}~\bibnamefont {Steinlechner}}, \bibinfo {author} {\bibfnamefont {J.}~\bibnamefont {Kofler}}, \bibinfo {author} {\bibfnamefont {J.-{\AA}.}\ \bibnamefont {Larsson}}, \bibinfo {author} {\bibfnamefont {C.}~\bibnamefont {Abell{\'a}n}}, \emph {et~al.},\ }\bibfield  {title} {\bibinfo {title} {Significant-loophole-free test of {Bell}'s theorem with entangled photons},\ }\href@noop {} {\bibfield  {journal} {\bibinfo  {journal} {Physical review letters}\ }\textbf {\bibinfo {volume} {115}},\ \bibinfo {pages} {250401} (\bibinfo {year} {2015})}\BibitemShut {NoStop}%
\bibitem [{\citenamefont {Rosenfeld}\ \emph {et~al.}(2017)\citenamefont {Rosenfeld}, \citenamefont {Burchardt}, \citenamefont {Garthoff}, \citenamefont {Redeker}, \citenamefont {Ortegel}, \citenamefont {Rau},\ and\ \citenamefont {Weinfurter}}]{bell4}%
  \BibitemOpen
  \bibfield  {author} {\bibinfo {author} {\bibfnamefont {W.}~\bibnamefont {Rosenfeld}}, \bibinfo {author} {\bibfnamefont {D.}~\bibnamefont {Burchardt}}, \bibinfo {author} {\bibfnamefont {R.}~\bibnamefont {Garthoff}}, \bibinfo {author} {\bibfnamefont {K.}~\bibnamefont {Redeker}}, \bibinfo {author} {\bibfnamefont {N.}~\bibnamefont {Ortegel}}, \bibinfo {author} {\bibfnamefont {M.}~\bibnamefont {Rau}},\ and\ \bibinfo {author} {\bibfnamefont {H.}~\bibnamefont {Weinfurter}},\ }\bibfield  {title} {\bibinfo {title} {Event-ready {Bell} test using entangled atoms simultaneously closing detection and locality loopholes},\ }\href@noop {} {\bibfield  {journal} {\bibinfo  {journal} {Physical review letters}\ }\textbf {\bibinfo {volume} {119}},\ \bibinfo {pages} {010402} (\bibinfo {year} {2017})}\BibitemShut {NoStop}%
\bibitem [{\citenamefont {Li}\ \emph {et~al.}(2018)\citenamefont {Li}, \citenamefont {Wu}, \citenamefont {Zhang}, \citenamefont {Liu}, \citenamefont {Bai}, \citenamefont {Liu}, \citenamefont {Zhang}, \citenamefont {Zhao}, \citenamefont {Li}, \citenamefont {Wang} \emph {et~al.}}]{bell5}%
  \BibitemOpen
  \bibfield  {author} {\bibinfo {author} {\bibfnamefont {M.-H.}\ \bibnamefont {Li}}, \bibinfo {author} {\bibfnamefont {C.}~\bibnamefont {Wu}}, \bibinfo {author} {\bibfnamefont {Y.}~\bibnamefont {Zhang}}, \bibinfo {author} {\bibfnamefont {W.-Z.}\ \bibnamefont {Liu}}, \bibinfo {author} {\bibfnamefont {B.}~\bibnamefont {Bai}}, \bibinfo {author} {\bibfnamefont {Y.}~\bibnamefont {Liu}}, \bibinfo {author} {\bibfnamefont {W.}~\bibnamefont {Zhang}}, \bibinfo {author} {\bibfnamefont {Q.}~\bibnamefont {Zhao}}, \bibinfo {author} {\bibfnamefont {H.}~\bibnamefont {Li}}, \bibinfo {author} {\bibfnamefont {Z.}~\bibnamefont {Wang}}, \emph {et~al.},\ }\bibfield  {title} {\bibinfo {title} {Test of local realism into the past without detection and locality loopholes},\ }\href@noop {} {\bibfield  {journal} {\bibinfo  {journal} {Physical review letters}\ }\textbf {\bibinfo {volume} {121}},\ \bibinfo {pages} {080404} (\bibinfo {year} {2018})}\BibitemShut {NoStop}%
\bibitem [{\citenamefont {Pironio}\ \emph {et~al.}(2010)\citenamefont {Pironio}, \citenamefont {Ac{\'\i}n}, \citenamefont {Massar}, \citenamefont {de~La~Giroday}, \citenamefont {Matsukevich}, \citenamefont {Maunz}, \citenamefont {Olmschenk}, \citenamefont {Hayes}, \citenamefont {Luo}, \citenamefont {Manning} \emph {et~al.}}]{pironio2010random}%
  \BibitemOpen
  \bibfield  {author} {\bibinfo {author} {\bibfnamefont {S.}~\bibnamefont {Pironio}}, \bibinfo {author} {\bibfnamefont {A.}~\bibnamefont {Ac{\'\i}n}}, \bibinfo {author} {\bibfnamefont {S.}~\bibnamefont {Massar}}, \bibinfo {author} {\bibfnamefont {A.~B.}\ \bibnamefont {de~La~Giroday}}, \bibinfo {author} {\bibfnamefont {D.~N.}\ \bibnamefont {Matsukevich}}, \bibinfo {author} {\bibfnamefont {P.}~\bibnamefont {Maunz}}, \bibinfo {author} {\bibfnamefont {S.}~\bibnamefont {Olmschenk}}, \bibinfo {author} {\bibfnamefont {D.}~\bibnamefont {Hayes}}, \bibinfo {author} {\bibfnamefont {L.}~\bibnamefont {Luo}}, \bibinfo {author} {\bibfnamefont {T.~A.}\ \bibnamefont {Manning}}, \emph {et~al.},\ }\bibfield  {title} {\bibinfo {title} {Random numbers certified by {B}ell’s theorem},\ }\href@noop {} {\bibfield  {journal} {\bibinfo  {journal} {Nature}\ }\textbf {\bibinfo {volume} {464}},\ \bibinfo {pages} {1021} (\bibinfo {year} {2010})}\BibitemShut {NoStop}%
\bibitem [{\citenamefont {Ac{\'\i}n}\ and\ \citenamefont {Masanes}(2016)}]{acin2016certified}%
  \BibitemOpen
  \bibfield  {author} {\bibinfo {author} {\bibfnamefont {A.}~\bibnamefont {Ac{\'\i}n}}\ and\ \bibinfo {author} {\bibfnamefont {L.}~\bibnamefont {Masanes}},\ }\bibfield  {title} {\bibinfo {title} {Certified randomness in quantum physics},\ }\href@noop {} {\bibfield  {journal} {\bibinfo  {journal} {Nature}\ }\textbf {\bibinfo {volume} {540}},\ \bibinfo {pages} {213} (\bibinfo {year} {2016})}\BibitemShut {NoStop}%
\bibitem [{\citenamefont {Colbeck}\ and\ \citenamefont {Renner}(2012)}]{colbeck2012free}%
  \BibitemOpen
  \bibfield  {author} {\bibinfo {author} {\bibfnamefont {R.}~\bibnamefont {Colbeck}}\ and\ \bibinfo {author} {\bibfnamefont {R.}~\bibnamefont {Renner}},\ }\bibfield  {title} {\bibinfo {title} {Free randomness can be amplified},\ }\href@noop {} {\bibfield  {journal} {\bibinfo  {journal} {Nature Physics}\ }\textbf {\bibinfo {volume} {8}},\ \bibinfo {pages} {450} (\bibinfo {year} {2012})}\BibitemShut {NoStop}%
\bibitem [{\citenamefont {P{\"u}tz}\ \emph {et~al.}(2014)\citenamefont {P{\"u}tz}, \citenamefont {Rosset}, \citenamefont {Barnea}, \citenamefont {Liang},\ and\ \citenamefont {Gisin}}]{putz2014arbitrarily}%
  \BibitemOpen
  \bibfield  {author} {\bibinfo {author} {\bibfnamefont {G.}~\bibnamefont {P{\"u}tz}}, \bibinfo {author} {\bibfnamefont {D.}~\bibnamefont {Rosset}}, \bibinfo {author} {\bibfnamefont {T.~J.}\ \bibnamefont {Barnea}}, \bibinfo {author} {\bibfnamefont {Y.-C.}\ \bibnamefont {Liang}},\ and\ \bibinfo {author} {\bibfnamefont {N.}~\bibnamefont {Gisin}},\ }\bibfield  {title} {\bibinfo {title} {Arbitrarily small amount of measurement independence is sufficient to manifest quantum nonlocality},\ }\href@noop {} {\bibfield  {journal} {\bibinfo  {journal} {Physical review letters}\ }\textbf {\bibinfo {volume} {113}},\ \bibinfo {pages} {190402} (\bibinfo {year} {2014})}\BibitemShut {NoStop}%
\bibitem [{\citenamefont {Earman}(2008)}]{earman2008pruning}%
  \BibitemOpen
  \bibfield  {author} {\bibinfo {author} {\bibfnamefont {J.}~\bibnamefont {Earman}},\ }\bibfield  {title} {\bibinfo {title} {Pruning some branches from “branching spacetimes”},\ }\href@noop {} {\bibfield  {journal} {\bibinfo  {journal} {Philosophy and Foundations of Physics}\ }\textbf {\bibinfo {volume} {4}},\ \bibinfo {pages} {187} (\bibinfo {year} {2008})}\BibitemShut {NoStop}%
\bibitem [{\citenamefont {Norton}(2003)}]{norton2003causation}%
  \BibitemOpen
  \bibfield  {author} {\bibinfo {author} {\bibfnamefont {J.}~\bibnamefont {Norton}},\ }\bibfield  {title} {\bibinfo {title} {Causation as folk science},\ }\href@noop {} {\bibfield  {journal} {\bibinfo  {journal} {Philosophia Mathematica}\ }\textbf {\bibinfo {volume} {3}},\ \bibinfo {pages} {1} (\bibinfo {year} {2003})}\BibitemShut {NoStop}%
\bibitem [{\citenamefont {Born}(2012)}]{born2012physics}%
  \BibitemOpen
  \bibfield  {author} {\bibinfo {author} {\bibfnamefont {M.}~\bibnamefont {Born}},\ }\href@noop {} {\emph {\bibinfo {title} {Physics in my generation}}}\ (\bibinfo  {publisher} {Springer Science \& Business Media},\ \bibinfo {year} {2012})\BibitemShut {NoStop}%
\bibitem [{\citenamefont {Reichenbach}(2012)}]{reichenbach2012philosophy}%
  \BibitemOpen
  \bibfield  {author} {\bibinfo {author} {\bibfnamefont {H.}~\bibnamefont {Reichenbach}},\ }\href@noop {} {\emph {\bibinfo {title} {The philosophy of space and time}}}\ (\bibinfo  {publisher} {Courier Corporation},\ \bibinfo {year} {2012})\BibitemShut {NoStop}%
\bibitem [{\citenamefont {Gisin}(2020{\natexlab{b}})}]{gisin2020mathematical}%
  \BibitemOpen
  \bibfield  {author} {\bibinfo {author} {\bibfnamefont {N.}~\bibnamefont {Gisin}},\ }\bibfield  {title} {\bibinfo {title} {Mathematical languages shape our understanding of time in physics},\ }\href@noop {} {\bibfield  {journal} {\bibinfo  {journal} {Nature Physics}\ }\textbf {\bibinfo {volume} {16}},\ \bibinfo {pages} {114} (\bibinfo {year} {2020}{\natexlab{b}})}\BibitemShut {NoStop}%
\bibitem [{\citenamefont {Brouwer}(1907)}]{brouwer1907dissertation}%
  \BibitemOpen
  \bibfield  {author} {\bibinfo {author} {\bibfnamefont {L.~E.~J.}\ \bibnamefont {Brouwer}},\ }\emph {\bibinfo {title} {Over de Grondslagen der Wiskunde}},\ \href@noop {} {Ph.D. thesis},\ \bibinfo  {school} {Universiteit van Amsterdam} (\bibinfo {year} {1907}),\ \bibinfo {note} {translated in~\cite[pp.13--101]{brouwer1975collected}}\BibitemShut {NoStop}%
\bibitem [{\citenamefont {Brouwer}(1912)}]{brouwer1912intuitionism}%
  \BibitemOpen
  \bibfield  {author} {\bibinfo {author} {\bibfnamefont {L.~E.~J.}\ \bibnamefont {Brouwer}},\ }\href@noop {} {\bibinfo {title} {{Intu{\"i}tionisme en Formalisme}}} (\bibinfo {year} {1912}),\ \bibinfo {note} {also in Wiskundig tijdschrift, 9, 1913. Translated in~\cite[pp.123--137]{brouwer1975collected}}\BibitemShut {NoStop}%
\bibitem [{\citenamefont {Copeland}(2012)}]{Copeland2012-COPTCT}%
  \BibitemOpen
  \bibfield  {author} {\bibinfo {author} {\bibfnamefont {B.~J.}\ \bibnamefont {Copeland}},\ }\bibfield  {title} {\bibinfo {title} {The {C}hurch-{T}uring thesis},\ }in\ \href@noop {} {\emph {\bibinfo {booktitle} {Stanford Encyclopedia of Philosophy}}},\ \bibinfo {editor} {edited by\ \bibinfo {editor} {\bibfnamefont {E.}~\bibnamefont {Zalta}}}\ (\bibinfo  {publisher} {Stanford Encyclopedia of Philosophy},\ \bibinfo {year} {2012})\BibitemShut {NoStop}%
\bibitem [{\citenamefont {Brouwer}(1913)}]{brouwer1913intuitionism}%
  \BibitemOpen
  \bibfield  {author} {\bibinfo {author} {\bibfnamefont {L.~E.~J.}\ \bibnamefont {Brouwer}},\ }\bibfield  {title} {\bibinfo {title} {Intuitionism and formalism},\ }\href@noop {} {\bibfield  {journal} {\bibinfo  {journal} {Bulletin of the American Mathematical Society}\ }\textbf {\bibinfo {volume} {20}},\ \bibinfo {pages} {81} (\bibinfo {year} {1913})}\BibitemShut {NoStop}%
\bibitem [{\citenamefont {Brouwer}(1948)}]{brouwer1948consciousness}%
  \BibitemOpen
  \bibfield  {author} {\bibinfo {author} {\bibfnamefont {L.~E.~J.}\ \bibnamefont {Brouwer}},\ }\bibfield  {title} {\bibinfo {title} {Consciousness, philosophy, and mathematics},\ }\href@noop {} {\bibfield  {journal} {\bibinfo  {journal} {Proceedings of the 10th International Congress of Philosophy, Amsterdam}\ ,\ \bibinfo {pages} {1235}} (\bibinfo {year} {1948})},\ \bibinfo {note} {translated in~\cite[pp.480--494]{brouwer1975collected}}\BibitemShut {NoStop}%
\bibitem [{\citenamefont {Posy}(2020)}]{posy2020mathematical}%
  \BibitemOpen
  \bibfield  {author} {\bibinfo {author} {\bibfnamefont {C.~J.}\ \bibnamefont {Posy}},\ }\href {https://doi.org/10.1017/9781108674485} {\emph {\bibinfo {title} {Mathematical intuitionism}}}\ (\bibinfo  {publisher} {Cambridge University Press},\ \bibinfo {year} {2020})\BibitemShut {NoStop}%
\bibitem [{\citenamefont {Bentzen}(2025)}]{bentzen2025iep}%
  \BibitemOpen
  \bibfield  {author} {\bibinfo {author} {\bibfnamefont {B.}~\bibnamefont {Bentzen}},\ }\bibfield  {title} {\bibinfo {title} {Intuitionism in mathematics},\ }in\ \href {https://iep.utm.edu/intuitionism-math/} {\emph {\bibinfo {booktitle} {The {Internet} Encyclopedia of Philosophy}}},\ \bibinfo {editor} {edited by\ \bibinfo {editor} {\bibfnamefont {J.}~\bibnamefont {Fieser}}\ and\ \bibinfo {editor} {\bibfnamefont {B.}~\bibnamefont {Dowden}}}\ (\bibinfo {year} {2025})\BibitemShut {NoStop}%
\bibitem [{\citenamefont {Shapiro}(2000)}]{shapiro2000thinking}%
  \BibitemOpen
  \bibfield  {author} {\bibinfo {author} {\bibfnamefont {S.}~\bibnamefont {Shapiro}},\ }\href@noop {} {\emph {\bibinfo {title} {Thinking about mathematics: The philosophy of mathematics}}}\ (\bibinfo  {publisher} {Oxford University Press},\ \bibinfo {address} {Oxford},\ \bibinfo {year} {2000})\BibitemShut {NoStop}%
\bibitem [{\citenamefont {Bentzen}(2023)}]{bentzen202xtwoity}%
  \BibitemOpen
  \bibfield  {author} {\bibinfo {author} {\bibfnamefont {B.}~\bibnamefont {Bentzen}},\ }\bibfield  {title} {\bibinfo {title} {Brouwer's intuition of twoity and constructions in separable mathematics},\ }\href {https://doi.org/10.1080/01445340.2023.2210908} {\bibfield  {journal} {\bibinfo  {journal} {History and Philosophy of Logic}\ }\textbf {\bibinfo {volume} {45}},\ \bibinfo {pages} {341} (\bibinfo {year} {2023})}\BibitemShut {NoStop}%
\bibitem [{\citenamefont {Troelstra}(1977)}]{troelstra1977choice}%
  \BibitemOpen
  \bibfield  {author} {\bibinfo {author} {\bibfnamefont {A.~S.}\ \bibnamefont {Troelstra}},\ }\href@noop {} {\emph {\bibinfo {title} {Choice sequences: a chapter of intuitionistic mathematics}}}\ (\bibinfo  {publisher} {Clarendon Press Oxford},\ \bibinfo {year} {1977})\BibitemShut {NoStop}%
\bibitem [{\citenamefont {van Atten}(2018)}]{vanatten2018creating}%
  \BibitemOpen
  \bibfield  {author} {\bibinfo {author} {\bibfnamefont {M.}~\bibnamefont {van Atten}},\ }\bibfield  {title} {\bibinfo {title} {{The Creating Subject, the Brouwer--Kripke Schema, and infinite proofs}},\ }\href {https://doi.org/10.1016/j.indag.2018.06.005} {\bibfield  {journal} {\bibinfo  {journal} {Indagationes Mathematicae}\ }\textbf {\bibinfo {volume} {29}},\ \bibinfo {pages} {1565} (\bibinfo {year} {2018})}\BibitemShut {NoStop}%
\bibitem [{\citenamefont {van Dalen}(1999{\natexlab{a}})}]{vandalen1999counterexamples}%
  \BibitemOpen
  \bibfield  {author} {\bibinfo {author} {\bibfnamefont {D.}~\bibnamefont {van Dalen}},\ }\bibfield  {title} {\bibinfo {title} {From {B}rouwerian counter examples to the creating subject},\ }\href {https://doi.org/10.1023/A:1026411905257} {\bibfield  {journal} {\bibinfo  {journal} {Studia Logica}\ }\textbf {\bibinfo {volume} {62}},\ \bibinfo {pages} {305} (\bibinfo {year} {1999}{\natexlab{a}})}\BibitemShut {NoStop}%
\bibitem [{\citenamefont {Dummett}(1977)}]{dummett1977elements}%
  \BibitemOpen
  \bibfield  {author} {\bibinfo {author} {\bibfnamefont {M.}~\bibnamefont {Dummett}},\ }\href@noop {} {\emph {\bibinfo {title} {Elements of intuitionism}}}\ (\bibinfo  {publisher} {Oxford University Press},\ \bibinfo {year} {1977})\BibitemShut {NoStop}%
\bibitem [{\citenamefont {Brouwer}(1955)}]{brouwer1955effect}%
  \BibitemOpen
  \bibfield  {author} {\bibinfo {author} {\bibfnamefont {L.~E.~J.}\ \bibnamefont {Brouwer}},\ }\bibfield  {title} {\bibinfo {title} {The effect of intuitionism on classical algebra of logic},\ }\href@noop {} {\bibfield  {journal} {\bibinfo  {journal} {Proceedings of the Royal Irish Academy}\ }\textbf {\bibinfo {volume} {57}},\ \bibinfo {pages} {113} (\bibinfo {year} {1955})},\ \bibinfo {note} {reprinted in~\cite[pp.551--554]{brouwer1975collected}}\BibitemShut {NoStop}%
\bibitem [{\citenamefont {Heyting}(1956)}]{heyting1956intuitionism}%
  \BibitemOpen
  \bibfield  {author} {\bibinfo {author} {\bibfnamefont {A.}~\bibnamefont {Heyting}},\ }\href@noop {} {\emph {\bibinfo {title} {Intuitionism: an introduction}}}\ (\bibinfo  {publisher} {North-Holland},\ \bibinfo {year} {1956})\BibitemShut {NoStop}%
\bibitem [{\citenamefont {Troelstra}\ and\ \citenamefont {van Dalen}(1988)}]{troelstravandalen1988constructivism}%
  \BibitemOpen
  \bibfield  {author} {\bibinfo {author} {\bibfnamefont {A.~S.}\ \bibnamefont {Troelstra}}\ and\ \bibinfo {author} {\bibfnamefont {D.}~\bibnamefont {van Dalen}},\ }\href@noop {} {\emph {\bibinfo {title} {Constructivism in mathematics. Vol. I}}},\ \bibinfo {series} {Studies in Logic and the Foundations of Mathematics}, Vol.\ \bibinfo {volume} {121}\ (\bibinfo  {publisher} {North-Holland},\ \bibinfo {address} {Amsterdam},\ \bibinfo {year} {1988})\BibitemShut {NoStop}%
\bibitem [{\citenamefont {Putnam}(1969)}]{putnam1969logic}%
  \BibitemOpen
  \bibfield  {author} {\bibinfo {author} {\bibfnamefont {H.}~\bibnamefont {Putnam}},\ }\bibinfo {title} {Is logic empirical?},\ in\ \href {https://doi.org/10.1007/978-94-010-3381-7_5} {\emph {\bibinfo {booktitle} {Boston Studies in the Philosophy of Science: Proceedings of the Boston Colloquium for the Philosophy of Science 1966/1968}}},\ \bibinfo {editor} {edited by\ \bibinfo {editor} {\bibfnamefont {R.~S.}\ \bibnamefont {Cohen}}\ and\ \bibinfo {editor} {\bibfnamefont {M.~W.}\ \bibnamefont {Wartofsky}}}\ (\bibinfo  {publisher} {Springer Netherlands},\ \bibinfo {address} {Dordrecht},\ \bibinfo {year} {1969})\ pp.\ \bibinfo {pages} {216--241}\BibitemShut {NoStop}%
\bibitem [{\citenamefont {Placek}(1999)}]{placek1999mathematical}%
  \BibitemOpen
  \bibfield  {author} {\bibinfo {author} {\bibfnamefont {T.}~\bibnamefont {Placek}},\ }\href@noop {} {\emph {\bibinfo {title} {Mathematical intuitionism and intersubjectivity: A critical exposition of arguments for intuitionism}}},\ Vol.\ \bibinfo {volume} {279}\ (\bibinfo  {publisher} {Springer Science \& Business Media},\ \bibinfo {year} {1999})\BibitemShut {NoStop}%
\bibitem [{\citenamefont {van Atten}(2004)}]{vanatten2004brouwer}%
  \BibitemOpen
  \bibfield  {author} {\bibinfo {author} {\bibfnamefont {M.}~\bibnamefont {van Atten}},\ }\href@noop {} {\emph {\bibinfo {title} {On Brouwer}}}\ (\bibinfo  {publisher} {Wadsworth/Thomson Learning},\ \bibinfo {address} {Belmont, CA},\ \bibinfo {year} {2004})\BibitemShut {NoStop}%
\bibitem [{\citenamefont {van Dalen}(1999{\natexlab{b}})}]{vandalen1999role}%
  \BibitemOpen
  \bibfield  {author} {\bibinfo {author} {\bibfnamefont {D.}~\bibnamefont {van Dalen}},\ }\bibfield  {title} {\bibinfo {title} {{The role of language and logic in Brouwer's work}},\ }in\ \href@noop {} {\emph {\bibinfo {booktitle} {Logic in Action. E}}},\ \bibinfo {editor} {edited by\ \bibinfo {editor} {\bibfnamefont {E.}~\bibnamefont {Orlovska}}}\ (\bibinfo  {publisher} {Springer},\ \bibinfo {address} {Vienna},\ \bibinfo {year} {1999})\ pp.\ \bibinfo {pages} {3--14}\BibitemShut {NoStop}%
\bibitem [{\citenamefont {Colyvan}(2024)}]{sep-mathphil-indis}%
  \BibitemOpen
  \bibfield  {author} {\bibinfo {author} {\bibfnamefont {M.}~\bibnamefont {Colyvan}},\ }\bibfield  {title} {\bibinfo {title} {{Indispensability Arguments in the Philosophy of Mathematics}},\ }in\ \href@noop {} {\emph {\bibinfo {booktitle} {The {Stanford} Encyclopedia of Philosophy}}},\ \bibinfo {editor} {edited by\ \bibinfo {editor} {\bibfnamefont {E.~N.}\ \bibnamefont {Zalta}}\ and\ \bibinfo {editor} {\bibfnamefont {U.}~\bibnamefont {Nodelman}}}\ (\bibinfo  {publisher} {Metaphysics Research Lab, Stanford University},\ \bibinfo {year} {2024})\ \bibinfo {edition} {{S}ummer 2024}\ ed.\BibitemShut {Stop}%
\bibitem [{\citenamefont {Bishop}(1967)}]{bishop1967foundations}%
  \BibitemOpen
  \bibfield  {author} {\bibinfo {author} {\bibfnamefont {E.}~\bibnamefont {Bishop}},\ }\href@noop {} {\emph {\bibinfo {title} {Foundations of Constructive Analysis}}}\ (\bibinfo  {publisher} {Academic Press},\ \bibinfo {address} {New York},\ \bibinfo {year} {1967})\BibitemShut {NoStop}%
\bibitem [{\citenamefont {Dummett}(1975)}]{dummett1975philosophical}%
  \BibitemOpen
  \bibfield  {author} {\bibinfo {author} {\bibfnamefont {M.}~\bibnamefont {Dummett}},\ }\bibfield  {title} {\bibinfo {title} {The philosophical basis of intuitionistic logic},\ }\href@noop {} {\bibfield  {journal} {\bibinfo  {journal} {Studies in Logic and the Foundations of Mathematics}\ }\textbf {\bibinfo {volume} {80}},\ \bibinfo {pages} {5} (\bibinfo {year} {1975})}\BibitemShut {NoStop}%
\bibitem [{\citenamefont {Martin-L{\"o}f}(2014)}]{martin2014truth}%
  \BibitemOpen
  \bibfield  {author} {\bibinfo {author} {\bibfnamefont {P.}~\bibnamefont {Martin-L{\"o}f}},\ }\href {https://pml.flu.cas.cz/uploads/PML-Leiden04Feb14.pdf} {\bibinfo {title} {Truth of empirical propositions}} (\bibinfo {year} {2014}),\ \bibinfo {note} {transcriptions of lectures given at Leiden University on February 4, 2014}\BibitemShut {NoStop}%
\bibitem [{\citenamefont {Raatikainen}(2004)}]{raatikainen2004conceptions}%
  \BibitemOpen
  \bibfield  {author} {\bibinfo {author} {\bibfnamefont {P.}~\bibnamefont {Raatikainen}},\ }\bibfield  {title} {\bibinfo {title} {Conceptions of truth in intuitionism},\ }\href@noop {} {\bibfield  {journal} {\bibinfo  {journal} {History and Philosophy of Logic}\ }\textbf {\bibinfo {volume} {25}},\ \bibinfo {pages} {131} (\bibinfo {year} {2004})}\BibitemShut {NoStop}%
\bibitem [{\citenamefont {Hansen}(2016)}]{hansen2016brouwer}%
  \BibitemOpen
  \bibfield  {author} {\bibinfo {author} {\bibfnamefont {C.~S.}\ \bibnamefont {Hansen}},\ }\bibfield  {title} {\bibinfo {title} {Brouwer's conception of truth},\ }\href@noop {} {\bibfield  {journal} {\bibinfo  {journal} {Philosophia Mathematica}\ }\textbf {\bibinfo {volume} {24}},\ \bibinfo {pages} {379} (\bibinfo {year} {2016})}\BibitemShut {NoStop}%
\end{thebibliography}%

\end{document}